%% file: rmt-markets.tex
\documentclass[showpacs]{revtex4}

\newcommand{\etal}{\textit{et al.} }

\usepackage{graphicx}
\usepackage{epsfig}
\usepackage{bm}

\catcode`\@=11

\def\eqbegin         {  \begin{eqnarray}  }
\def\eqend           {  \end{eqnarray}  }
\def\beq{\begin{equation}}
\def\eeq{\end{equation}}

\def\hs_2{\hspace{2mm}}
\def\hs_3{\hspace{3mm}}

\begin{document}
\title{Random Matrix Theory Analysis of Cross Correlations in Financial Markets}
\author{
Akihiko Utsugi$^1$, Kazusumi Ino$^2$, Masaki Oshikawa$^1$}

\affiliation{
$^1$ Department of Physics,Tokyo Institute of Technology, 
 Oh-okayama 2-12-1, Meguro-ku,  Tokyo,  152-8551, Japan \\
$^2$ Department of Pure and Applied Sciences, University of Tokyo,
Komaba 3-8-1, Meguro-ku,  Tokyo,  153-8902, Japan}

\thispagestyle{empty}

\begin{abstract} 
We confirm universal behaviors such as  eigenvalue distribution and 
spacings predicted by Random Matrix Theory (RMT) for 
the cross correlation matrix of the daily stock prices of 
Tokyo Stock Exchange from 1993 to 2001,  
which have been reported for New York Stock Exchange in previous studies. 
It is shown that  the random part of the eigenvalue distribution 
of the cross  correlation matrix is stable even when 
deterministic correlations are present.
Some deviations 
in the small eigenvalue statistics 
outside the bounds of the universality class of 
RMT are not completely explained
with the deterministic correlations as proposed in previous studies.
We study the effect of randomness on deterministic 
correlations and find that randomness causes  
a repulsion between deterministic eigenvalues and 
the random eigenvalues. This is  interpreted as a reminiscent of  
``level repulsion'' 
in RMT and explains some deviations from the previous studies 
observed in the market data.
We also study correlated groups of issues in these markets and 
propose a refined method to identify  correlated groups based on RMT.
Some characteristic differences between properties  of 
Tokyo Stock Exchange and New York Stock Exchange 
are found.

\end{abstract}

\maketitle
\section{introduction}
The price changes of  securities such as stocks involve 
various economic backgrounds as well as interaction between securities. 
They seem to be quite complicated. 
Conventionally financial economists  model 
 the price changes of securities  by stochastic processes 
({\it random walks}) \cite{merton}. It is a basic 
ingredient of modern portfolio theory \cite{port}. 
Although the use of stochastic processes is common in finance,
the validity of such a formulation should be  empirically 
tested e.g. by statistical properties of the markets, since 
the underlying ergodic property of a market may be hard to be 
established.

Recently the statistical characterizations of financial markets based 
on physics concepts and methods attract considerable attentions 
\cite{econo}. 
Given that a stochastic model is valid,  some statistical properties 
of  the market should be derived as outsets of stochasticity. For example, 
the cross correlation matrix among $N$ securities  
can be regarded as a random matrix and 
it may be legitimate to expect that it  shares 
universal properties of a corresponding ensemble of 
Random Matrix Theory (RMT) in an appropriate large $N$-limit 
(since $N$ is usually large). 
This has been  confirmed by several studies on actual stock markets 
\cite{laloux,plerou,plerou2}.  The bulk of the eigenvalue distribution of 
the cross correlation matrix of a major index (S\&P500) of 
New York Stock Exchange (NYSE) 
is found to follow the eigenvalue distribution 
of the Wishart matrix \cite{laloux}
,  which 
is a random correlation matrix constructed from mutually uncorrelated
time series \cite{baker, edelman}. 
Also the eigenvalue spacing statistics are found to follow those of the 
Gaussian Orthogonal Ensemble (GOE) \cite{plerou}.

The aim of this paper is to yield further supports 
on the applicability of RMT to analysis of stock markets. 

In Sec.\ref{sec2},  we give a brief review on the relevant results of RMT. 
We describe our data sample in Sec.\ref{sec3}.
In Sec.\ref{sec_universal},  
we test predictions of RMT for the  cross-correlation matrix 
for the daily prices of the issues in  
 Tokyo Stock Exchange (TSE) from 1993 to 2001. 
The quantities we calculated are the distribution of the 
eigenvalues,  the nearest and next-nearest neighbor spacings,  
rigidity and a certain moment of eigenvector components. 
We find a 
good agreement with the real data within the RMT bounds for 
the eigenvalues. Indeed there are clear deviations outside the bounds 
which indicate  the presence of deterministic correlations among issues. 
In Sec.\ref{sec_sta},  
we consider random walks with deterministic correlations and show that 
the bulk part of the eigenvalue distribution of the correlation matrix 
is stable. 
In Sec.\ref{levelrepul},  
 we closely examine the distribution of the moment
 of eigenvector component.
Eigenvectors corresponding to the eigenvalues outside the 
RMT bounds deviate from the RMT prediction.
According to Ref.~\cite{plerou2},  
the deviating eigenvalues at the lower edge are
a consequence of the strong correlations among a few issues. However,  
we find that the observed data 
is not explained quantitatively by this reasoning alone. Therefore we
analyze the effect of randomness on deterministic correlations between 
issues and find  an interplay between deterministic correlations 
and randomness. We argue that
it gives a refined explanation on the deviations.
In Sec.\ref{sec4},  we identify  groups of strongly correlated issues from
the information of the non-random eigenvectors. The ways of grouping in
TSE and NYSE show some differences.

\section{Brief Review on Random Matrix Theory}
\label{sec2}
\subsection{Wishart Matrix}
Let $S_i(t)$ be   a price at time $t$  of a stock labeled by $i$ 
$(i=1, 2, \cdots, N , t=1, 2, \cdots, T) $. 
The change of price at time $t$  can be measured by    
\begin{equation}
G_i(t)\equiv \ln S_i(t+1)-\ln S_i(t). 
\label{eq4}   
\end{equation}
Here we take logarithm of the prices because the fluctuation of 
stock prices is typically given by the geometric Brown motion. 
Since 
\begin{eqnarray}
G_i(t)
      &\simeq&\frac{S_i(t+1)-S_i(t)}{S_i(t)}
\label{eq7}
\end{eqnarray}
$G_i(t)$ is approximately the return of the issue $i$ from $t$ to $t+1$. 
We also define the normalized return $g_i(t)$ as follows. 
\begin{equation}
g_i(t)\equiv \frac{G_i(t)-\langle G_i\rangle_T}{\sigma_i}. 
\label{eq8}
\end{equation}
$\langle \cdots \rangle_T$ indicates the time series average of $T$ steps and 
the dispersion $\sigma_i$ is given by 
\begin{equation}
\sigma_i\equiv \sqrt{\langle G_i^2\rangle_T-\langle G_i\rangle_T^2}.
 \label{eq9}
\end{equation}
Then the correlation matrix $C$ is expressed in terms of $g_i(t)$ 
\begin{equation}
C_{ij}\equiv \langle g_ig_j \rangle _T.  
\label{eq10}
\end{equation}
$C$ is a real symmetric matrix with positive eigenvalues. 

We will model the price of stocks as a stochastic process (random walk).
For $N$ random walks $x_i(t), \hspace{2mm}(i=1, 2, \cdots, N)$,  
a matrix $M$ which is defined by $M_{ti} ={x_i(t)}$  is a $T\times N$ matrix. 
The cross correlation matrix $W$ is defined as follows  
\begin{equation}
W_{ij}\equiv \langle x_ix_j \rangle_T = \frac{1}{T}M^tM,   
\label{eq12}
\end{equation}
where $M^t$ is the transposition of $M$. 
A purely random case with a uniform dispersion $\sigma$ is given by 
\begin{eqnarray}
\langle x_i(t)\rangle =0,  \\
\langle x_i(t)x_j(\tau)\rangle =\sigma^2\delta_{ij}\delta_{t\tau}.
\label{eq11}
\end{eqnarray}
Here  $\langle \cdots \rangle $ indicates 
the average over the random variable phase space. 
In this case,  $W$ is called the Wishart matrix \cite{baker,edelman}.
We can include ``true''
correlations among issues by 
replacing $\delta_{ij}$ in (\ref{eq11}) by a non-diagonal matrix $\widetilde{C}$.  
We will call $\widetilde{C}$ as {\it deterministic correlation} 
while we call $C$ or $W$  as  {\it cross correlation}.

\subsection{Eigenvalue Statistics of Random Matrices} 

Let us  summarize the relevant results of RMT
to which we will refer in this paper. 

In the limit $N \rightarrow \infty ,  T \rightarrow \infty $ with 
$Q\equiv T/N$ fixed,   
the eigenvalue distribution $\rho(\lambda)$ for the
Wishart matrix becomes~\cite{sengupta}
\begin{equation}
\rho(\lambda)=\frac{Q}{2\pi\sigma^2}
\frac{\sqrt{(\lambda_{max}-\lambda)(\lambda-\lambda_{min})}}{\lambda}
\label{eq26}
\end{equation}
\begin{equation}
\lambda_{min}^{max}=\sigma^2(1+\frac{1}{Q}\pm2\sqrt{\frac{1}{Q}})
\label{eq27}
\end{equation}
(\ref{eq26}) is exact at $N \rightarrow \infty ,  T\rightarrow \infty$ with 
$Q\equiv T/N=const$. It is   approximately valid at finite $N$ and $T$ 
when $N$ and $T$ are not small.  
According to (\ref{eq26}) (\ref{eq27}),  
the eigenvalues of the Wishart matrix distribute only in the range $(\lambda_{min}, \lambda_{max})$.

Next we consider the Gaussian ensembles of random matrices. 
In the Gaussian ensembles,  
the probability of a matrix $H$ to be in the infinitesimal volume element 
 $dH$ ($dH$ is given by the product of infinitesimal of independent elements) 
is given by $P(H)dH$ where $P(H)$ 
\begin{equation}
P(H)=A\exp(-a\sum_i|\lambda_i|^2).
\label{eq13}
\end{equation}
Here $a$ is a parameter which characterizes the ensemble,   
$\lambda_i$ is the eigenvalue of $H$ and $A$ is the normalization 
constant. For general ensembles,  one replaces 
the term $\sum_i |\lambda_i|^2$  by $\sum_i V(\lambda_i)$ with 
a function $V(\lambda)$. 
For example,  one can add the quartic  or higher order terms,  but 
it is known that,  in the large $N$-limit ($N$ is the size of $H$),  
the model flows to the Gaussian model \cite{hikami}. 
The Gaussian models are classified by 
the symmetry of the  matrix as   
i) Gaussian Orthogonal Ensemble  (GOE), the ensemble invariant under 
the orthogonal group,  ii) Gaussiann Symplectic Ensemble (GSE), the ensemble invariant 
under the symplectic group,  
and iii) Gaussiann Unitary Ensemble (GUE),  the ensemble invariant under 
the unitary group.  
Since the correlation matrix $C$ is 
real symmetric,  the ensemble relevant to our analysis is GOE. 
For GOE,  the volume element $dH$ is given by 
\begin{equation}
dH=\prod_{i \leq j}dH_{ij}. 
\label{eq15}
\end{equation}

To obtain the statistical measure of the eigenvalue distribution 
$P(\lambda_1, \lambda_2, \cdots, \lambda_N)$,  
one expresses $H$ as  the product of 
the diagonal matrix with eigenvalue entries and the other variables,  
and then  
integrates the other variables.
In this way,  we get  the measure  
\begin{equation}
\prod_{i<j}|\lambda_i-\lambda_j|^\beta \prod_k d\lambda_k.
\label{eq16}
\end{equation}
Here $\beta =1$ for GOE,  $\beta=2$ for GUE and $\beta=4$ for GSE.
Thus the eigenvalue distribution for a Gaussian ensemble is determined 
by $\beta$.
By this way,  
we get the eigenvalue distribution for a general potential $V$ as follows.
\begin{eqnarray}
&&P(\lambda_1, \lambda_2, \cdots, \lambda_N)
=A'\exp[-\beta(\sum_{k=1}^N\frac{V(\lambda_k)}{\beta}-\sum_{i<j}\ln|\lambda_i-\lambda_j|)], 
\label{eq18}
\end{eqnarray}
where $A'$ is the normalization constant. 
From (\ref{eq18}),  one sees that 
the statistical properties at the short  spacing
 between eigenvalues  are dominated by $-{\rm ln}|\lambda_i-\lambda_j|$ 
and the total potential is negligible. 
Thus $\beta$ determines the eigenvalue spacing at short distance. 
For each $\beta$,  the level spacing has been closely studied \cite{mehta}.
As the correlation matrix is real symmetric,  we expect that 
its statistical properties of the eigenvalue spacing are given by 
$\beta=1$. 
One can characterize the statistical properties of eigenvalue spacing by 
the nearest neighbor spacing $P_{nn}$,  the next-nearest 
neighbor spacing $P_{nnn}$,  and the "rigidity" $\Delta(L)$. 
$P_{nn}$ and $P_{nnn}$ are for short-range correlations while 
$\Delta(L)$ is for long-range correlations.
$\Delta(L)$ is defined as 
\begin{equation}
\Delta(L) \equiv \frac{1}{L}\left<\min_{A, B}\int_{\lambda-\frac{L}{2}}^{\lambda+\frac{L}{2}}(F(\lambda')-A\lambda'-B)^2d\lambda'\right>_\lambda, 
\label{eq19}
\end{equation}
where $F(\lambda)$ is given by 
\begin{equation}
F(\lambda)=\sum_k\Theta(\lambda-\lambda_k)
\label{eq20}
\end{equation}
with the Heaviside function $\Theta$. $F(\lambda)$ counts 
the number of eigenvalues below $\lambda$.  
The meaning of $\Delta(L)$ is that one fits  $F(\lambda)$ 
by a line in an interval with a width $L$ around each eigenvalue,  
and take the average of the deviations of the fit. 
$\Delta(L)$  is small when the eigenvalue spacing has a uniform distribution.

For GOE,  $P_{nn}, P_{nnn}$ and $\Delta(L)$ are given by \cite{mehta}, 
\begin{eqnarray}
P_{nn}(s)&=&\frac{\pi s}{2}\exp(-\frac{\pi}{4}s^2) \label{eq21}
\\ 
P_{nnn}(s)&=&\frac{2^{18}}{3^6 \pi^3}s^4\exp(-\frac{64}{9\pi}s^2) \label{eq22} 
\\
\Delta(L)&=&\frac{1}{15}L^{-4} \int_0^L du(L-u)^3(2L^2-9Lu-3u^2) \nonumber 
\\ 
&\times& (\frac{1}{2}\delta(u)-Y(u)) 
\label{eq23}
\end{eqnarray}
$Y(u)$ is called 2-spectral cluster function given by 
\begin{equation}
Y(u)=\left(\frac{\sin(\pi u)}{\pi u}\right)^2+\frac{d}{du}\left(\frac{\sin(\pi u)}{\pi u}\right)\int_u^\infty\frac{\sin(\pi t)}{\pi t}dt.
\label{eq24}
\end{equation}

According to RMT,  the distribution of 
components of an eigenvector  
of GOE  is the normal distribution with mean $0$ and 
dispersion $N$.  A useful quantity in characterizing the distribution 
of components is the  Inverse Participation Ratio (IPR) 
\cite{mehta,fyodorov}.  For each eigenvector ${\boldmath u_k}$,  
IPR is defined by the following formula. 
\begin{equation}
I_k \equiv \sum_{i=1}^N u_{ki}^4,  
\label{eq51}
\end{equation}
where $u_{ki}$ is the $i$-th component of ${\boldmath u_k}$. 
For example,  let us consider the case  $u_{ki}$ is $1/\sqrt{L}$ 
for $1 \leq i \leq L$  
and 0 for the other $i$'s. This gives  $I_k =1/L$. Thus IPR can be 
interpreted as the inverse of the number of components which 
differ from zero significantly. 
In RMT,  the expectation value of IPR is  
\begin{equation}
\langle I_k \rangle =N\int_{-\infty}^{\infty} u_{ki}^4\frac{1}{\sqrt{2\pi N}}\exp\left(-\frac{u_{ki}^2}{2N}\right)du_{ki}=\frac{3}{N}. 
\label{eq52}
\end{equation}

\section{Market Data}
\label{sec3}
 The data we analyzed are daily stock prices of i) Tokyo Stock Exchange (TSE) from 
1993 January to 2001 June and 
 ii) S\&P 500 index of New York Stock Exchange (NYSE) 
from 1991 January to 2001 July.
As for S\&P,  the daily price data for a different period has been analyzed by 
Laloux \etal~\cite{laloux}. Also the 30 minute price data for NYSE has been 
studied by Plerou \etal~\cite{plerou,plerou2}.   
In the TSE data, the number of data points (the days that the market is open)
is 1848.
We analyze, among all the issues in TSE, 
the 493 issues which are traded in all of the 1848 days.
We select the data of these issues and analyze it. 
For this data,  $N=493$ and $T=1848$.  
In the S\&P500 data,  the number of data points is 2599. 
We select  the issues which have been selected in S\&P500 index 
before 1991 and analyze their prices. 
They are amount to  297. For this data,  $N=297$ and $T=2598$.

\section{Universal Random Properties of 
Cross Correlations In Stock Markets}
\label{sec_universal}
In  Refs. \cite{laloux,plerou},    
the cross correlation matrices of  NYSE data are analyzed 
and  found that they exhibit remarkable agreement with 
the predictions of universality properties of 
RMT for the small eigenvalues' distribution,  
their nearest and next-nearest neighbor  spacings,  rigidity 
and IPR. 
In this section, we perform a similar analysis on the TSE data and 
confirm these properties.  We also use the S\&P data for comparison.

\begin{figure}
\epsfxsize =11cm
\centerline{\epsfbox{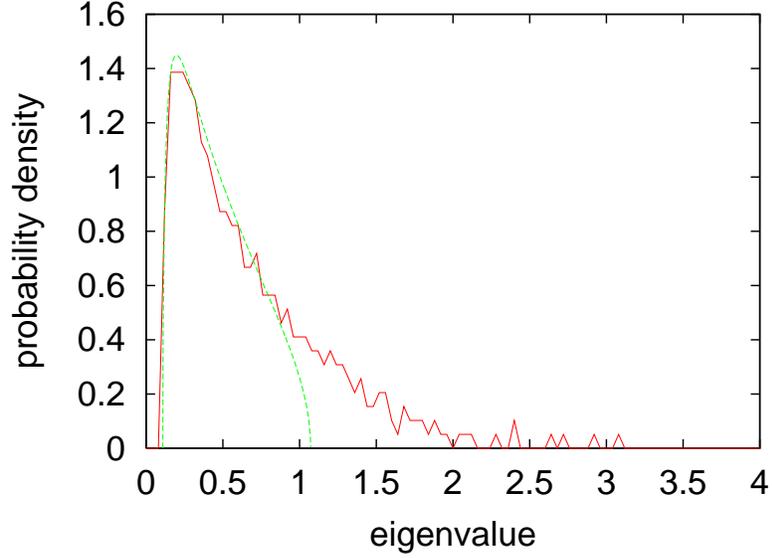}}
\caption{The  figure 
shows the eigenvalue distribution for the correlation 
matrix of TSE. 
The  line in each figure is for the real data and the dotted line is 
for the Wishart matrix.  
We use  (\ref{eq26}) multiplied by   $N'/N$  for fitting where 
$N'$ is the number of eigenvalues within $[\lambda_{min}, \lambda_{max}]$.
$\sigma^2$ is fitted to the optimized value by the least square method. 
$\sigma^2 =0.47 (0.53)$ for TSE (S\&P).
For TSE (S\&P), a Kolmogorov-Smirnov test  in the 
fitted region cannot reject the hypothesis that 
the RMT prediction is the correct description at 
the 30\% (60\%) confidence level.}
\label{fig3}
\end{figure}

We diagonalize the correlation matrices of TSE and S\&P, to obtain 
the eigenvalues and
the eigenvectors $\mbox{\boldmath $u$}_k$ $(k=1, \cdots, N)$. 
$k$ is smaller for a large eigenvalue. 
 For TSE,  $\sigma^2=1$ and $Q=N/T=3.75$ give 
$\lambda_{min}=0.23$ and $\lambda_{max}=2.30$,  also for S\&P,  $Q=8.75$ 
gives $\lambda_{min}=0.43$ and $\lambda_{max}=1.79$. 
We  fit the distributions 
by optimizing $\sigma^2$ smaller than 1, 
as discussed in Ref.~\cite{laloux}. 
Fig.\ref{fig3} shows the eigenvalue distribution for TSE.
We see that 
the small eigenvalue distribution of the correlation matrix 
of TSE is well reproduced
by RMT.  There are large eigenvalues beyond the bound 
$[\lambda_{\rm min}, \lambda_{\rm max}]$ predicted by the Wishart matrix. 
The largest eigenvalue we obtain is 121.6 (52.2) for TSE (S\&P) and 
is interpreted as the factor for market trend as readily verified by 
examining the corresponding eigenvector.
The multitude of this factor 
to  the price changes of individual stocks is given by $\lambda_1/N$,  which
 is 0.247 (0.176) for TSE (S\&P). Thus TSE is more correlated 
with the trend factor than S\&P.

\begin{figure}
\epsfxsize =5cm
\centerline{\epsfbox{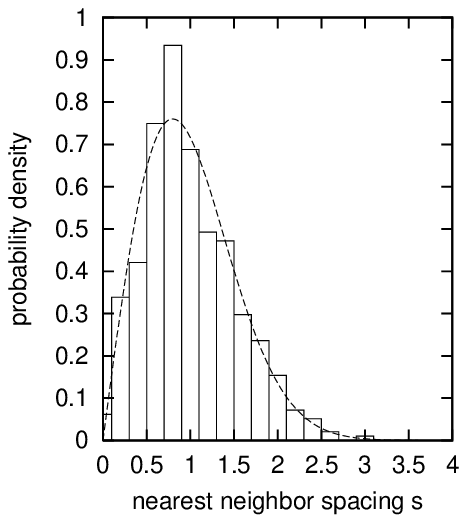}}
\epsfxsize =5cm
\centerline{\epsfbox{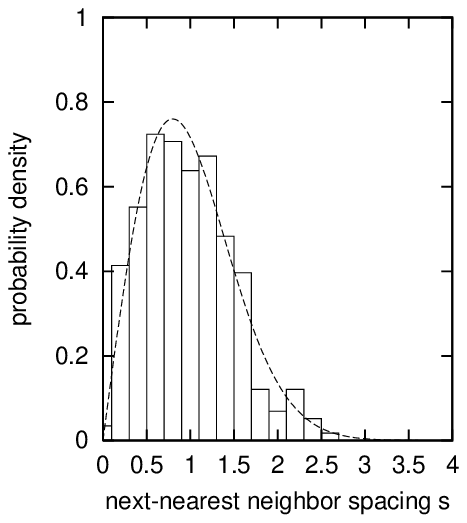}}
  \caption{
The  figures are the nearest  and 
the next-nearest neighbor spacing distribution for TSE  
compared to the prediction of RMT indicated by the dotted line. 
A Kolmogorov-Smirnov test 
cannot reject the hypothesis that the GOE prediction 
is the correct description at the 30\% (80\%) confidence level for 
the nearest neighbor spacing for  TSE (S\&P),  
at the 80\% (60\%) confidence level for the next-nearest neighbor spacing
for TSE (S\&P). }
\label{fig6}
\end{figure}

\begin{figure}
\epsfxsize =8cm
\centerline{\epsfbox{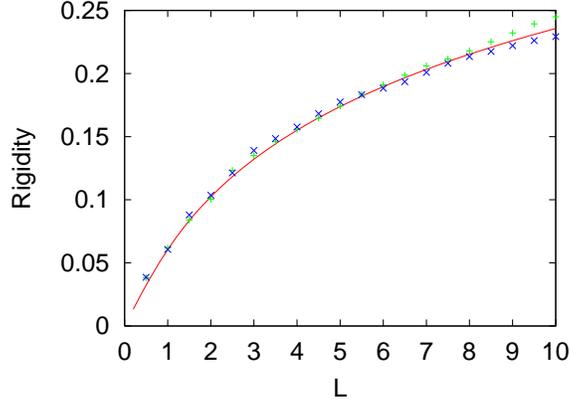}}
  \caption{
The plus mark is the rigidity $\Delta(L)$ for TSE while 
the  $\times$ mark is the rigidity for S\&P.
The  line is the prediction of RMT.
A Kolmogorov-Smirnov test cannot reject the hypothesis that the 
GOE prediction is the correct description at the 
80\% confidence level both for TSE and S\&P. }
\label{fig7}
\end{figure}

\begin{figure}
 \epsfxsize =6cm
\centerline{\epsfbox{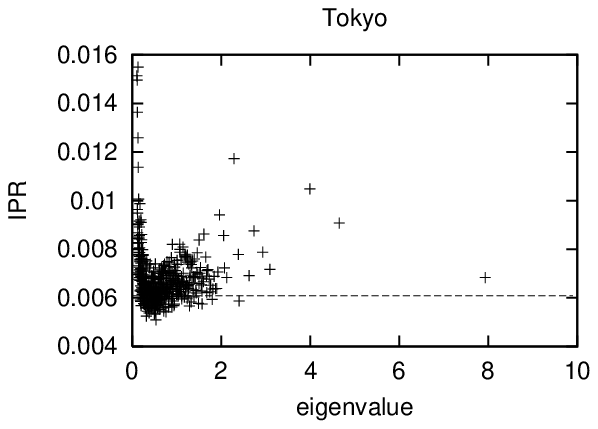}}
 \epsfxsize =6cm
\centerline{\epsfbox{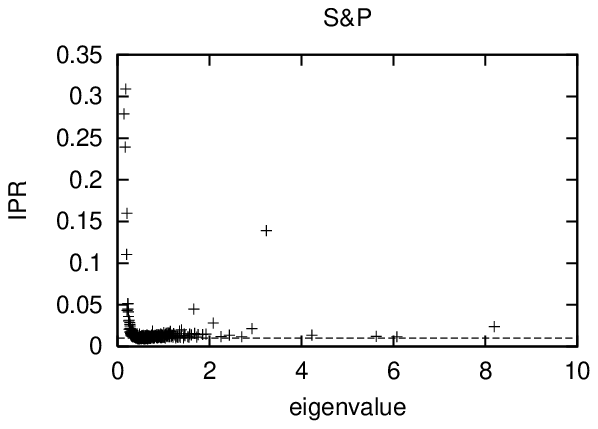}}
 \epsfxsize =6cm
\centerline{\epsfbox{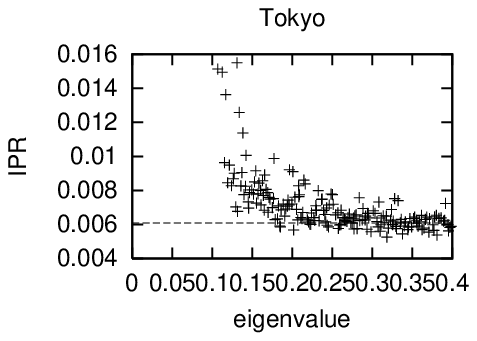}}
 \epsfxsize =6cm
\centerline{\epsfbox{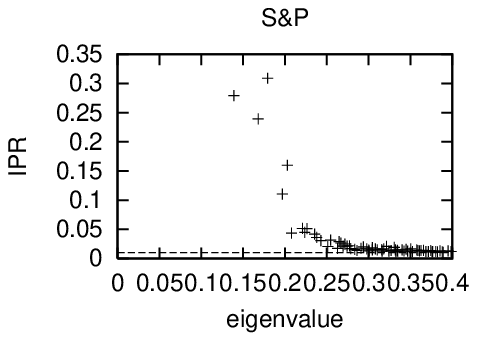}}
   \caption{The upper  two figures are IPR for TSE and S\&P. 
The lower two figures are IPR for TSE and S\&P 
at small eigenvalues.
The dotted lines are the prediction of RMT. 
}
\label{fig8}
 \end{figure}

Next we compare  
spacings of the nearest neighbor and the 
next-nearest neighbor eigenvalues,  
and the rigidity with the predictions of RMT.
To examine the statistics
 of the eigenvalue spacing,  we first do the ``unfolding'' 
transformation on the data.  
The ``unfolding" transformation is described in  \cite{plerou2}. 
After doing  the ``unfolding'' transformation on the eigenvalues 
 below $\lambda_{\rm max}$,
we compare their nearest-neighbor and 
next-nearest neighbor spacing distributions 
to the ones for GOE. The theoretical predictions for 
the nearest neighbor spacing and the next-nearest neighbor spacing
are given in (\ref{eq21}) and (\ref{eq22}) respectively. 
We show in Fig.~\ref{fig6} the spacings of small eigenvalues for TSE. 
It shows a good agreement with 
the prediction of RMT. For the rigidity $\Delta(L)$,  
the theoretical prediction is  given in eq.~(\ref{eq23}). 
The rigidity of the eigenvalues of the cross correlation matrix 
for TSE  below $\lambda_{\rm max}$  is compared to RMT 
in Fig.\ref{fig7}. Fig.\ref{fig7} shows that 
the rigidity agrees well with the prediction of RMT.

In Fig.\ref{fig8},  we plot the 
calculated IPR for the eigenvectors of the cross 
correlation matrix of TSE. 
One sees that 
IPR agrees with the prediction of RMT around $1$.
There are also eigenvectors whose IPR are larger than the RMT 
prediction.  
These eigenvalues are from deterministic correlations. 
As in Fig.\ref{fig8},  such  deviations can be seen at the large eigenvalues.
However one also sees that 
there is a deviation in small eigenvalues. 
This deviation is 
concentrated at the lower edge. 
A simple model was constructed by Plerou \etal~\cite{plerou2}.
We will study  this deviation  closely  in Sec.\ref{levelrepul}.

As mentioned,  we also performed the same analysis on the S\&P data
for comparison.
Results for rigidity and IPR are shown  in  figs. \ref{fig7} and \ref{fig8}. 
We found that  the conclusions  of Plerou \etal~\cite{plerou,plerou2} 
for 30 minutes data of NYSE on eigenvalue spacings 
also hold   for our daily S\&P data.

\newpage

\section{Stability of 
Eigenvalue Distribution of the Wishart Matrix 
 In The Presence of Deterministic Correlations }
\label{sec_sta}
In the previous section,  we found that the small eigenvalue distributions 
of the cross correlation matrices of TSE and S\&P   
are reproduced well by the ones 
of the Wishart matrix,  as previously found in Ref.~\cite{laloux}.  
The Wishart matrix is generated by the  random walks 
without any deterministic correlations while 
the real stock data has a   distribution 
of large eigenvalues, showing a deviation from the Wishart matrix. 
This indicates  the existence of deterministic correlations. 

Thus, in this section, we examine the stability of the
random eigenvalue distribution of 
the cross correlation matrix $W$ of random walks 
when one includes  deterministic correlations. 

Let us consider a set of random walks whose deterministic
correlation matrix has 
a finite number of large eigenvalues  and other eigenvalues are small.
We assume that $T \times N$ matrix $\{M_{ti}=x_i(t)\}$ has a 
deterministic correlation of the form 
\begin{equation}
\langle M_{ti} \rangle =0,   
\label{eq39}
\end{equation}
\begin{equation}
\langle M_{ti}M_{\tau j}\rangle =D_{t\tau}\widetilde{C}_{ij}.
\label{eq40}
\end{equation}
The cross correlation matrix at step $T$ is given by $M^{t}M$. 
As in RMT,  the eigenvalue distribution of $M^tM$ is calculated 
from the Green function,  
\begin{equation}
G(\lambda)\equiv \left< \frac{1}{\lambda-M^t M}\right > , 
\label{eq41}
\end{equation}
by the formula 
\begin{equation}
\rho(\lambda)=\frac{1}{2\pi N}\lim_{\epsilon\to0}{\rm Im}[{\rm Tr}G(\lambda-i\epsilon)-{\rm Tr}G(\lambda+i\epsilon)].
\label{eq43}
\end{equation}
The present case was studied in Ref.~\cite{sengupta}.
Using the replica method,  
a Dyson-type equation for $G$ was obtained at $N, T \rightarrow 
\infty$ with $Q=T/N$ fixed as follows
\begin{equation}
G(\lambda)=\frac{1}{\lambda-\widetilde{C}{\rm Tr}\left(\frac{D}{1-D{\rm Tr}(\widetilde{C}G(\lambda))}\right)}.
\label{eq42}
\end{equation}
$(\ref{eq26})$ is readily obtained by putting $\widetilde{C}=\sigma^2{\bf 1}$, $D={\bf 1}/T$ and taking the trace of (\ref{eq42})
\begin{equation}
{\rm Tr}G(\lambda)=\frac{N}{\lambda-\sigma^2\frac{1}{1-\frac{\sigma^2}{T}{\rm Tr}G(\lambda)}}.
\label{eq44}
\end{equation}
Solving this second-order algebraic equation for 
${\rm Tr}G(\lambda)$ and putting
 the solution to (\ref{eq43}) yields (\ref{eq26}) and (\ref{eq27}).

Now we assume that $\widetilde{C}$ has $L$ large eigenvalues 
$\lambda_k^{\widetilde{C}}$
$(k=1, 2, \cdots, L)$ and the other $N-L$ 
eigenvalues $\lambda^{\widetilde{C}}_k (k=L+1, \cdots, N)$. 
We set $\lambda^{\widetilde{C}}_k (k=L+1, \cdots, N)$ to be 
a  same value $\lambda^{\widetilde{C}}_s$. 
Since the trace of 
the cross correlation matrix equals $N$ by definition,   we have 
\begin{equation}
\lambda_s^{\widetilde{C}}=\frac{N-\sum_{k=1}^L\lambda_k^{\widetilde{C}}}{N-L}.
\label{eq45}
\end{equation}
We also assume no temporal correlations thus set  
$D={\bf 1}/T$.


From (\ref{eq42}), the eigenvalues $\lambda^{G}_k(\lambda)$ of 
$G(\lambda)$ are given by
\begin{equation}
\lambda_k^G(\lambda)=\frac{1}{\lambda-\lambda_k^{\widetilde{C}}\frac{1}{1-\frac{1}{T}({\rm Tr}_{S}\widetilde{C}G + {\rm Tr}_{L}\widetilde{C}G)}}.
\label{eq46.0}
\end{equation}
Let us split the trace as ${\rm Tr}={\rm Tr}_{L}+{\rm Tr}_{S}.$ 
Here ${\rm Tr}_{L}$ and ${\rm Tr}_{S}$ are the trace
 over the eigenspace spanned by 
the eigenvectors for $\lambda^{G}_{k} (k=1, \cdots L),  
\lambda^{G}_{k} (k=L+1, \cdots N) 
$ respectively. 
Summation over $k=L+1, \cdots, N $ gives 
\begin{equation}
{\rm Tr}_{S}\lambda_k^G(\lambda)=\frac{N-L}{\lambda-\lambda_s^{\widetilde{C}}\frac{1}{1-\frac{1}{T}({\rm Tr}_{S}\widetilde{C}G + {\rm Tr}_{L}\widetilde{C}G)}}.
\label{eq46.1}
\end{equation}
For $N$ large,  $\rho(\lambda)$ should have finite supports around 
$\lambda^{\widetilde{C}}_k$ in the real axis of $\lambda$. 
We denote supports for large and small 
eigenvalues $D_S$ and $D_L$.
We assume that the case 
\begin{equation}
\lambda^{\widetilde{C}}_{s} \ll \lambda^{\widetilde{C}}_k,  (k=1, \cdots L), 
\label{cond1}
\end{equation}  
when $D_S$ and $D_L$ don't have an overlap. In that case,  
$\lambda^{G}_k$ $(k=1, \cdots, L)$ is analytic 
in $D_S$ while 
$\lambda^{G}_k$ $(k=L+1, \cdots, N)$ has a branch cut.    
Thus in $D_S$,  
$\rho(\lambda)$ is determined by the imaginary part of ${\rm Tr}_{S}G$. 
For ${\rm Tr}_{S}G$,  the contribution from    $\lambda^{G}_k  (k=1, \cdots,  L)$ 
comes from
 the right hand of (\ref{eq46.1}). 
Since $\lambda^{G}_k (k=1, \cdots L)$ is analytic in the neighborhood of $D_S$,  
${\rm Tr}_L G$ is bounded by a constant. Since $\lambda^{G}_k$ is an 
algebraic function of $N$ and the  scaling behavior  
consistent with (\ref{eq46.0}) is $O(1)$,   the constant 
can be taken to be   independent of $N$.         
Thus if for $k = 1, \cdots,  L$ 
\begin{equation} 
L \lambda^{\widetilde{C}}_k \ll N\lambda^{\widetilde{C}}_{s}, 
\label{cond2}
\end{equation}  
then  
\[{\rm Tr}_{S}\widetilde{C}G= \lambda^{\widetilde{C}}_s{\rm Tr}_S G 
\gg {\rm Tr}_{L}\widetilde{C}G \] 
 for $N$ large because ${\rm Tr}_S G$ gets large as $N \rightarrow \infty$.  Then 
(\ref{eq46.1}) is approximated by  
\begin{equation}
{\rm Tr}_SG(\lambda)=
\frac{N-L}{\lambda-\lambda^{\widetilde{C}}_s\frac{1}{1-\frac{\lambda_s^{\widetilde{C}}}{T}
{\rm Tr}_S G(\lambda)}}.
\label{eq49}
\end{equation}
(\ref{eq49}) is equal to (\ref{eq44}) when 
$\sigma^2=\lambda_s^{\widetilde{C}}$ and 
$N$ is replaced by $N-L$. By putting the solution of (\ref{eq49}) to (\ref{eq43}),  we get 
\begin{equation}
\rho(\lambda) \simeq \frac{N-L}{N}\frac{Q}{2\pi{\lambda^{\widetilde{C}}_s}}
\frac{\sqrt{(\lambda_{\rm max}-\lambda)(\lambda-\lambda_{\rm min})}}{\lambda}.
\label{eq50}
\end{equation}
This formula is valid under (\ref{cond1}) and (\ref{cond2}).  
Note that there is a trade-off
 between $N, L, \lambda^{\widetilde{C}}_s, \lambda^{\widetilde{C}}_{k} (k=1, \cdots,  L)$ under (\ref{cond1}) and (\ref{cond2}).  
Thus $N-L$ eigenvalue distribution of this model can be approximated by 
the one for the Wishart matrix. 

\begin{figure}
\epsfxsize =6cm
\centerline{\epsfbox{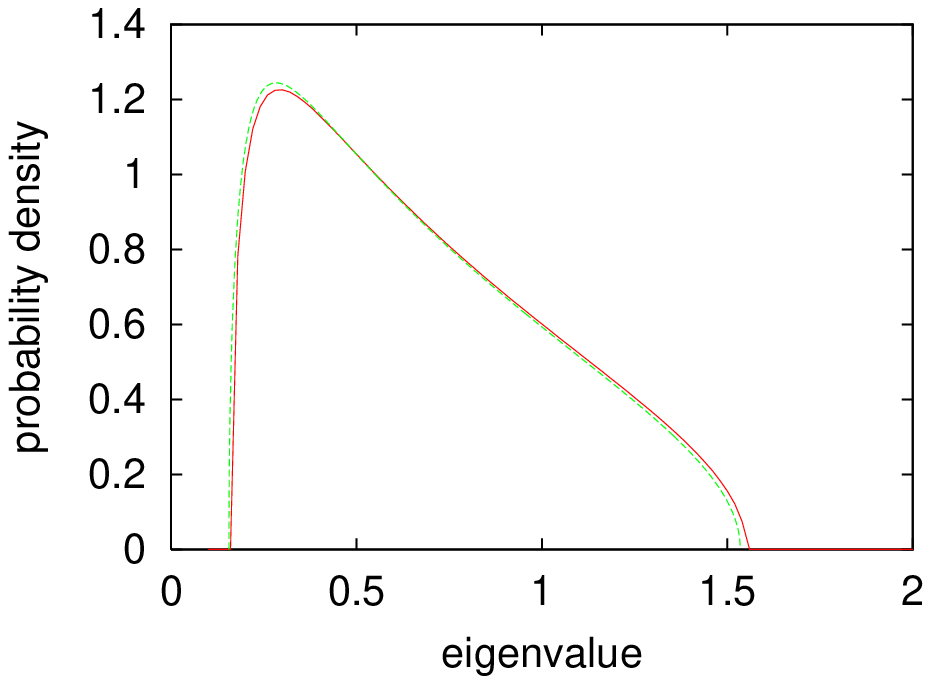}}
\centerline{\epsfbox{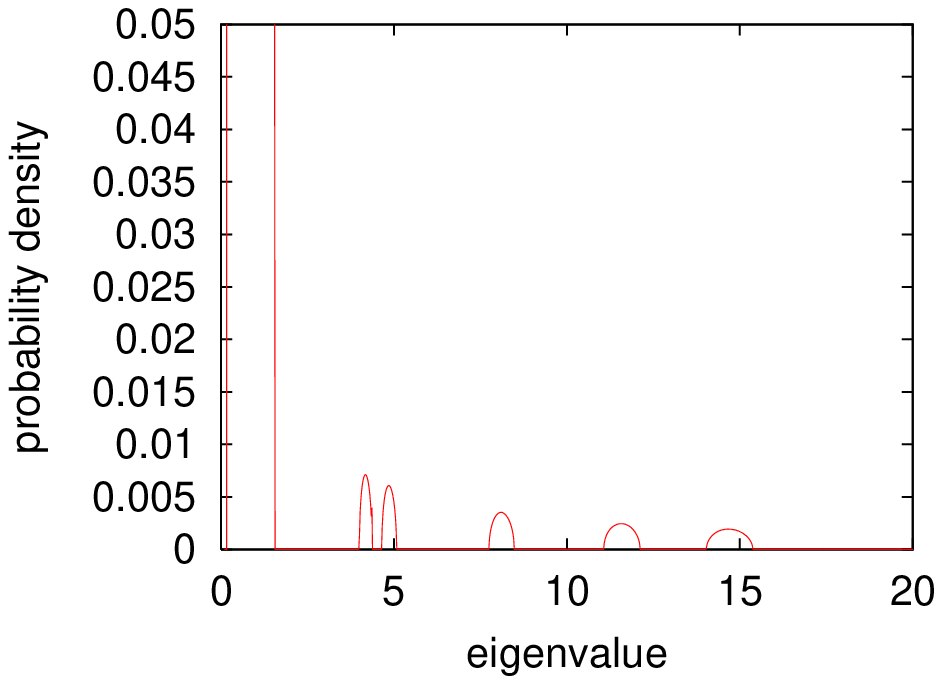}}
\centerline{\epsfbox{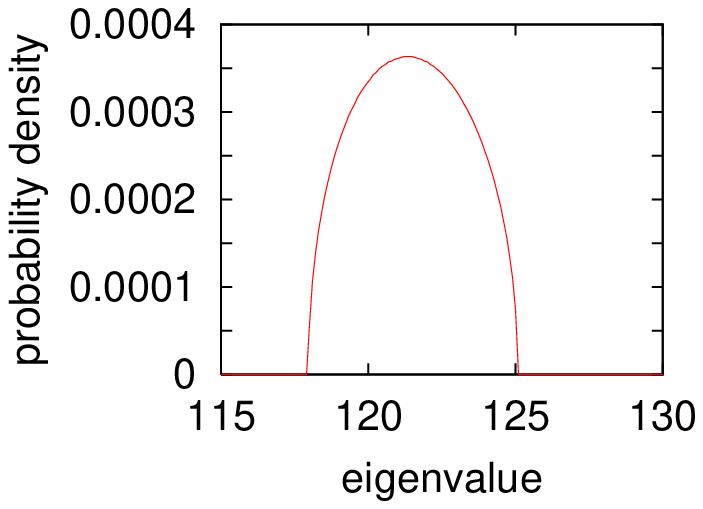}}
\caption{The line is for the model with large eigenvalues 
of the real correlation matrix 
while the dotted line is for (\ref{eq26}) with $\sigma^2 =\lambda^{\widetilde C}_s$ 
in (\ref{eq45}). 
The small eigenvalue distribution (the upper graph) is very close. 
The middle graphs are the large eigenvalue distribution.
We take the large eigenvalues of the real correlation matrix 
as 121.6, 14.5, 11.4, 7.9, 4.7, 4.0 which are 
found for TSE. 
The eigenvalues are observed in neighborhood of these values.
Also,  the  observed eigenvalues  have a finite width by 
the effect of randomness. The width of the observed eigenvalues is 
wider for the larger eigenvalues.
}
\label{fig5}
 \end{figure}

To conclude, the distribution of the small eigenvalues remains the same in the 
$N \rightarrow \infty$, as long as the 
numbers of the large eigenvalues of the deterministic correlation 
$\widetilde{C}$ is finite and they appear only outside of $D_S$.

To confirm the validity of the approximation, we performed 
a Monte-Carlo simulation with 6 large eigenvalues. We choose the 
large eigenvalues to be 121.6,  14.5, 11.4,  7.9,  4.7,  4.0 which 
are the observed large eigenvalues of TSE. 
The result is shown in Fig.\ref{fig5}. We see that the large eigenvalues 
correspond to the large eigenvalues of the real correlation matrix while 
the small eigenvalue distribution is well reproduced by the one for 
the Wishart matrix.  We also examined other 
values of large eigenvalues and obtained similar results. 
Moreover,  the probability of observed eigenvalues  has a finite width by 
the effect of  randomness. The width of an observed eigenvalue is 
wider for a larger eigenvalue. 

\newpage 
\section {Level Repulsion of Deterministic Correlations by  
Randomness}
\label{levelrepul}
According to Plerou \etal~\cite{plerou}, the deviation at small eigenvalues 
arises from  strong correlations among a small number of issues. 
This is illustrated well by the following model.
We consider a model that $N$ issues have an equal correlation $c$:
\begin{eqnarray}
\widetilde{C}= \left( 
\begin{array}{cccc} 
1 & c & \cdots & c \\
c & 1 & & \vdots \\ 
\vdots & & \ddots & c \\ 
c & \cdots & c & 1\\ 
\end{array} \right) 
\label{eq53} 
\end{eqnarray} 
$\widetilde{C}$ has  an eigenvalue $1+(N-1)c$ with no degeneracy and 
an eigenvalue $1-c$ with degeneracy $N-1$.
The eigenvalue $1-c$ becomes small if $c$ 
is close to 1 i.e. strong correlation. Its eigenvectors have 
non-zero components at the correlated issues,  resulting in 
a large IPR.

However this reasoning of large IPR eigenvectors at small eigenvalues 
 is not sufficient to 
 explain two facts. Firstly,  eigenvectors with large IPR 
appear only {\it below} the bulk of the eigenvalue distribution of the 
Wishart matrix,  concentrating at the lower edge. 
Since the correlation $c$ should be distributed in a wide range,  
 eigenvectors with large IPR should also be distributed in a wide range.
Thus the absence of small eigenvalues with large IPR within the bulk is 
puzzling.
Secondly,  each eigenvector with large IPR is observed at a smaller value 
than expected from the model above. As the largest non-diagonal element
of the correlation matrix of TSE (S\&P)  is 0.74 (0.83),  
eq.~(\ref{eq53}) tells that the eigenvector with large IPR with the smallest 
eigenvalue should be observed at 0.26 (0.18). 
Actually the smallest eigenvalue 
with large IPR is observed at 0.11 (0.14) 
which is smaller than  
the lower bound of the eigenvalue distribution of the  Wishart matrix.

These two facts motivate us to study  
the interplay between deterministic correlations and randomness.
We consider a model of random walks with 
a deterministic correlation matrix $\widetilde{C}$,  and examine IPR of 
eigenvectors of 
 the cross  correlation matrix $C$. As a simple model, 
we assume $\widetilde{C}$ to have a following form : 
\begin{eqnarray}
\widetilde{C}=
\left(\begin{array}{ccccc}
\widetilde{C}_1 & 0 &\cdots &\cdots & 0 \\
0 & \widetilde{C}_2 & & & \vdots \\
\vdots & & \ddots &  \\
\vdots& & & \widetilde{C}_L &0 \\
0 & \cdots & \cdots & 0 & {\boldmath 1}\\
\end{array}
\right).
\label{eq54}
\end{eqnarray}
Here $\widetilde{C}_l~(l=1, \cdots, L)$ and ${\bf 1}$ 
are  
\begin{eqnarray} 
\widetilde{C}_l&=&
\left(
\begin{array}{cccc}
1 & c_l & \cdots & c_l \\
c_l & 1 & & \vdots \\
\vdots & & \ddots & c_l \\
c_l & \cdots & c_l & 1 \\
\end{array}
\right)  \hspace{1 cm}
{\bf 1}=\left(\begin{array}{cccc}
1 & 0 & \cdots & 0 \\
0 & 1 & & \vdots \\
\vdots & & \ddots & 0 \\
0 & \cdots & 0 & 1 \\
\end{array}
\right).
\label{eq55}
\end{eqnarray}
The form of $\widetilde{C}$  assumes $L$ groups of 
 issues with strong correlations. 
We consider $N$ random walks $x_i(t)$ with 
\begin{equation}
\langle x_i(t)x_j(\tau)\rangle =\widetilde{C}_{ij}\delta_{t\tau}
\label{eq56}
\end{equation} 
and examine their $T$-step cross correlation matrix   
\begin{equation}
C_{ij}=\frac{1}{T}M^{t}M=\frac{1}{T}\sum_{t=1}^Tx_i(t)x_j(t). 
\label{eq57}
\end{equation}

\begin{figure}
 \epsfxsize =8cm
\centerline{\epsfbox{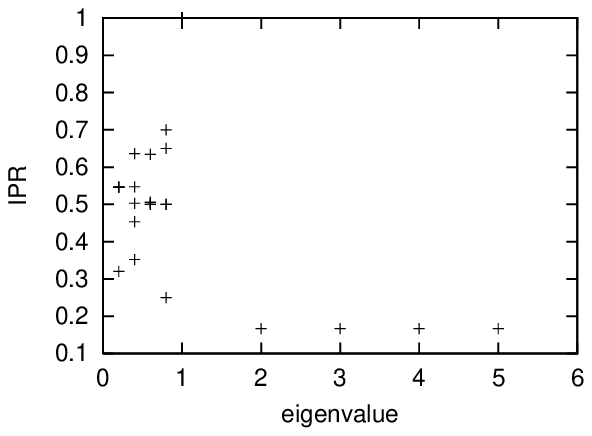}}
 \epsfxsize =8cm 
\centerline{\epsfbox{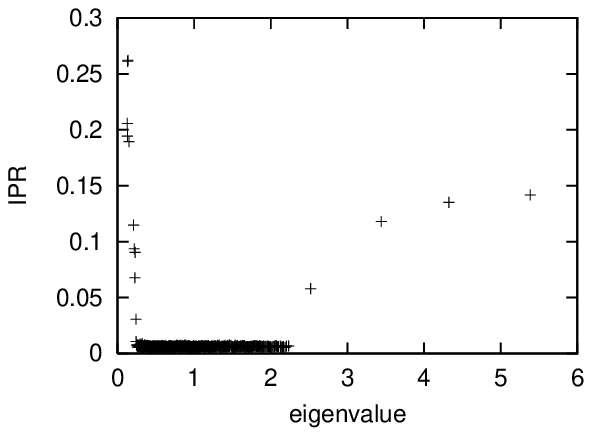}}
    \caption{
The upper graph is the IPR of the eigenvectors of the
real correlation matrix  $\widetilde{C}$ given by 
(\ref{eq54}-\ref{eq57}). The lower graph is the IPR for the eigenvector of $C$.
In the simulation,  we set $
N=493,  T=1847,  M=6,  L=4,  c_1=0.8, c_2=0.6, c_3=0.4$ and $c_4=0.2$.}
\label{fig9}
 \end{figure}

\begin{figure}
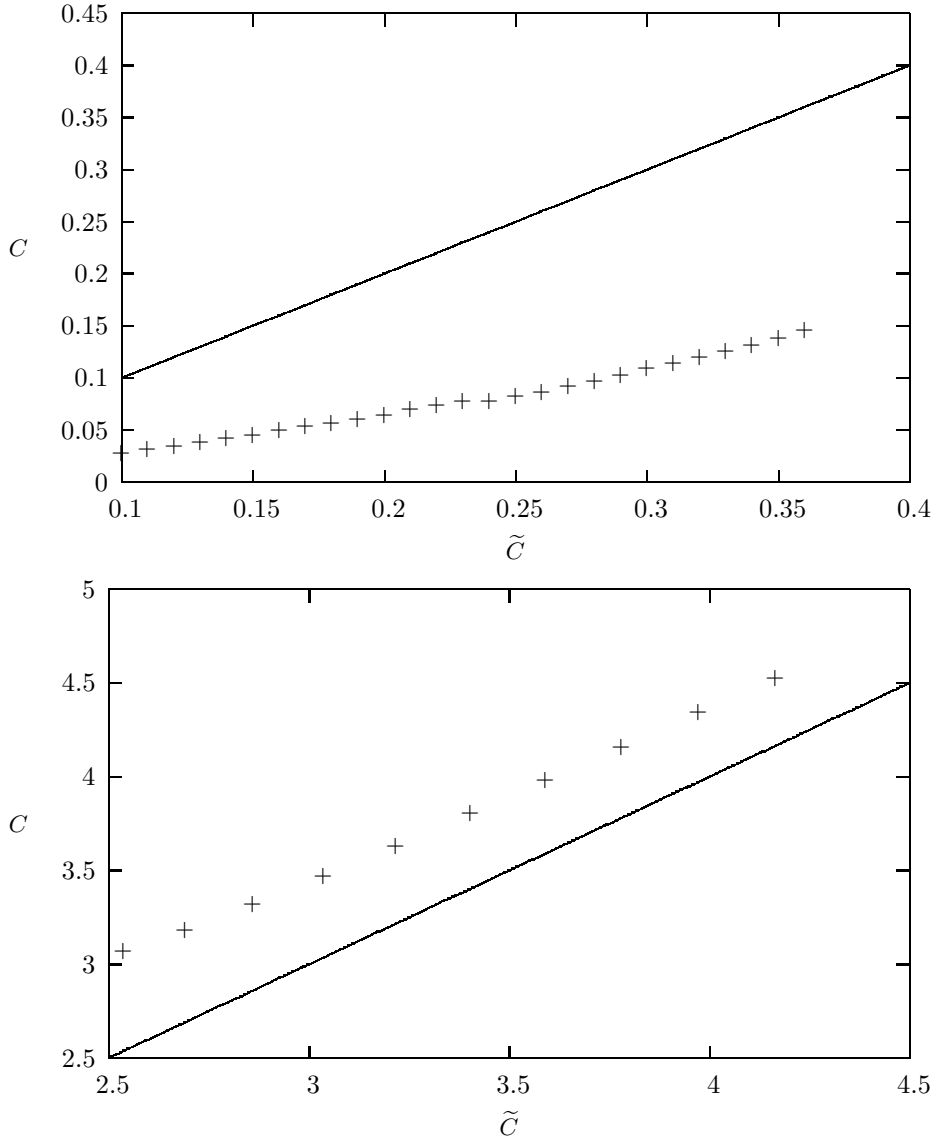

\include{graph10-1n}
\include{graph10-2n}
  \caption{The effect of level repulsion on the eigenvalues of $C$. 
The horizontal axis is the small (large) eigenvalue of $\widetilde{C}$ 
and the vertical axis is the corresponding eigenvalue of $C$. 
The upper (lower) graph is for  the case where eigenvalues of $\widetilde{C}$ 
are smaller (larger) than 1.
The crosses are the result of a Monte-Carlo simulation based on eqs.
(\ref{eq56}) and 
(\ref{eq57}). The straight line corresponds to the absence of the effect of 
randomness, when the eigenvalues of $C$ are identical to those of 
$\widetilde{C}$.
The eigenvalues of $C$ are repelled from the bulk vicinity of $1$.}
\label{fig10}
\end{figure}

We set $N=493$ and $T=1847$ following our TSE data. We set the number $L$ of 
strongly correlated groups to be $4$ and the number of issues $M$ 
participating each group to be $6$. We choose the correlations 
to be $c_1=0.8, c_2=0.6, c_3=0.4, c_4=0.2$. 
We performed a Monte Carlo simulation of this model. We present IPR of 
the eigenvalues in Fig.\ref{fig9}.  
Fig.\ref{fig9} shows that eigenvalues with large IPR distribute 
outside the bounds of eigenvalue distribution from randomness 
as in the real stock data.
In this model,  there should be 20 (counting degeneracies) 
small eigenvalues with large IPR in the simple model above,  but the observed 
ones with large IPR only amount to 10. 
This implies that, {\em when 
small eigenvalues arising from a strong correlation appear within 
the bounds of the Wishart matrix,  IPRs of 
their eigenvectors get smaller and 
cannot be distinguished from the random eigenvalues}.
This is one effect of randomness on deterministic correlations.
We also note that even for the eigenvectors which have 
larger IPR than the RMT value,  their IPRs are smaller than expected.

Moreover  $\widetilde{C}$ 
has small eigenvalues 0.2 and 0.4 while  
 the corresponding eigenvalues of $C$ distribute  
 in the vicinity of  0.14 and 0.22 respectively. 
On the other hand,  the eigenvalues of $C$ corresponding to 
the large eigenvalues of $\widetilde{C}$  are shifted to  
larger values than the original values.
Namely {\it the eigenvalues of $C$ from the deterministic 
correlation are repelled from the  distribution 
of  the random eigenvalues}. 
We performed Monte Carlo simulations by changing the parameters 
for $\widetilde{C}$ and got similar results. 
This may be interpreted as 
a  manifestation of the universal effect of randomness,  
called ``level repulsion''   \cite{potter}.  According to RMT,  the eigenvalues 
of random matrices are repelled from each other by 
the logarithmic potential in $-\ln|\lambda_i-\lambda_j|$ in (\ref{eq16}). 
Even when some deterministic terms are present,  this logarithmic 
potential  causes a repulsion between eigenvalues. 
This universal effect
 has been observed for various systems such as 
levels of complicated nuclei.
In the present case,  
deterministic correlations between random walks are repelled from 
the bulk distribution of  the random eigenvalues. The eigenvalues in 
the RMT bounds form a repulsive  potential and it  
 repels the eigenvalues outside them.

We can  
deduce this  ``level repulsion''  by solving the Dyson-type equations
 (\ref{eq39}-\ref{eq43}) numerically. 
We assume for simplicity that the eigenvalues of 
 $\widetilde{C}$ are $1$ except one eigenvalue smaller or larger than 1. 
We solve (\ref{eq39}-\ref{eq43}) numerically for $N=293$ and $T=1847$ 
and obtain the relation 
between the smaller (or larger) eigenvalue and the corresponding eigenvalue 
of $C$.   The result is shown in Fig.\ref{fig10}.
Fig.\ref{fig10} shows that smaller (larger) 
eigenvalues of $\widetilde{C}$ are repelled by the bulk distribution around $1$ and 
are observed as  smaller (larger) eigenvalues of $C$.

Thus we found two interplays between deterministic correlations 
and randomness.  
Namely,  when groups of issues have strong correlations,  
 it results in large and small eigenvalues in the cross correlation matrix. 
Some of these eigenvalues are soaked up within the RMT 
bounds and their IPR becomes as small as the RMT value. 
They cannot be distinguished from random  eigenvalues.
On the other hand,   eigenvalues from deterministic 
correlations outside the RMT bounds feel the repulsive potential 
generated by 
 the bulk distribution of randomness. At the lower edge,  
they are shifted to  smaller values.
We believe that  these  give the explanation for 
two deviations we raised in this section. 

\newpage 
\section{Groups of Issues Formed by Strong Correlation}
\label{sec4}
We have seen that the existence of 
 a group of issues with strong correlation results in 
eigenvalues of the cross correlation matrix  with large IPR. 
Conversely,  by examining the eigenvectors with large IPR, 
 we may identify groups formed by strong correlations. 

\begin{figure}
  \epsfxsize=8cm
 \centerline{\epsfbox{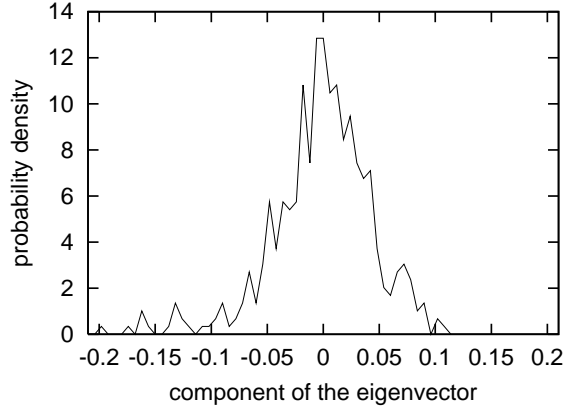}}
   \caption{
The component distribution of the eigenvector for 
the six-th largest eigenvalue $4.0$ of TSE. 
The components distribute continuously and it is hard 
to distinguish  the components from correlations. 
}
\label{fig11}
 \end{figure}

For NYSE,  Plerou et al  \cite{plerou} examined the eigenvectors of 
large eigenvalues and distinguished  strongly correlated issues 
by  a criteria to have  a 
large component in these eigenvectors.
They found that the groups are formed according to the industrial sectors.
However we found a difficulty to apply their method to TSE.   
Because eigenvectors for large eigenvalues have significant components 
not only from correlations but also from randomness,  
even if an issue has a large component in an eigenvector of large eigenvalue,  
it is difficult to tell whether it is from the effect of deterministic 
correlation or 
just from randomness.
Especially for TSE, the effect of deterministic correlations is
apparently not strong enough to make the separation straightforward.
As we examined the eigenvectors of the large eigenvalues,
we found it impossible to separate the group of strongly
correlated issues. 
For example,  Fig.~\ref{fig11} shows the component distribution of 
the eigenvector for the six-th largest eigenvalue 4.0 in TSE. 
One sees that the components have a continuous distribution and it is hard 
to separate large components due to deterministic correlaitons.

Therefore, here we propose a supplementary method to identify strongly
correlated components. As we saw in 
Sec.~\ref{levelrepul},
when a group of issues is formed by  strong correlations, 
they  not only  have  
a large component in the eigenvectors of the corresponding large eigenvalue,  
but also have a large component 
in the eigenvectors of the corresponding small eigenvalue. 
On the other hand,  
issues which do not have strong correlations with others should 
have the normal distribution in eigenvectors.
Namely, the deviation from the normal distribution indicates the issue 
is correlated with others. 
To quantify how an issue has a distribution different from  the normal distribution, 
we define a  quantity $Z_i$ as follows. 
\begin{equation}
Z_i=\sum_{k:I_k \geq \delta_{\rm th}}u_{ki}^2, 
\label{eq58}
\end{equation}
where $\delta_{\rm th}$ is a threshold for IPR. 
$Z_i$ is the sum of the square of 
$i$-th component of the eigenvectors which have IPR $\geq \delta_{\rm th}$. 
We set $\delta_{\rm th}=0.008(0.02)$ for TSE (S\&P),  which sort out 
41(28) eigenvectors. 
If $i$-th issue has no true correlation with others, the components 
$u_{ki}$ of the eigenvectors follow the normal distribution, and hence 
the probability of having a large $Z_i$ should be small. Thus the $i$-th 
issue may be regarded as significantly correlated if 
$Z_i$ is larger than a certain threshold $\alpha_{\rm th}$. 
We choose $\alpha_{\rm th}$ so that the probability of 
$Z_i \geq \alpha_{\rm th}$ is 1 \% if the eigenvector components 
for the $i$-th issue follows the normal distribution. 
For our data,  $\alpha_{\rm th}$ =0.131 (0.162)
 for TSE (S\&P).
If $i$-th issue has a large component in an eigenvector,   
we consider it to be in the corresponding  group of strong correlations when 
$Z_i \geq \alpha_{\rm th}$. 

We applied this method to large eigenvalues observed in 
our market data. The results are  
shown in {\bf TABLE 1-2}.
\\  
\\
{\bf TABLE 1} TSE issues with $Z_i \geq \alpha_{\rm th}$ 
\begin{tabbing}
xxxxxxxxxxxxx \= xxxxxxxxxx \= xxxxxxxxxxxxxxxxxxxxxxxxxxxxxx \=xxxxxxxxxxxxxxx \kill \\
{\it Eigenvector}\> {\it TSE code}  \>{\it Company Name}  \>{\it Sector}   \\
\>  \>  \>   \\
\(u_{2}\) \> 6701 \>NEC  \>Electric Products   \\
\(u_{2}\) \> 6702 \>Fujitsu  \> Electric Products \\
\(u_{2}\) \> 8035  \>Tokyo Electron  \> Electric Products \\
 \>  \>  \>  \\
\(u_{3}\) \> 1888 \>Wakachiku Construction  \>Construction  \\
\(u_{3}\) \> 8834 \>Douwa Real Estate  \>Real Estate  \\
 \>  \>  \>  \\
\(u_{4}\) \>9501  \>Tokyo Electric Power  \>Electric Power  \\
\(u_{4}\) \>9503  \>Kansai Electric Power  \>Electric Power  \\
\(u_{4}\) \>9504  \>Chuugoku Electric Power  \>Electric Power  \\
\(u_{4}\) \>9506  \>Tohoku Electric Power  \>Electric Power  \\
\(u_{4}\) \>9509  \>Hokkaido Electric Power  \>Electric Power  \\
 \>  \>  \>  \\

\(u_{5}\) \> 1888 \>Wakachiku Construction  \>Construction  \\
\(u_{5}\) \> 8834 \>Douwa Real Estate  \>Real Estate  \\
\(u_{5}\) \> 1801 \> Taisei Corporation \> Construction  \\
\(u_{5}\) \> 1804 \>Satou Kogyo  \>Construction   \\
\(u_{5}\) \> 1805 \>Tobishima Construction  \> Construction  \\
\(u_{5}\) \> 1806 \>Fujita Corporation  \> Construction  \\
\(u_{5}\) \> 1886 \>Aoki Corporation  \>Construction   \\
\(u_{5}\) \> 8601 \>Daiwa Securities  \> Finance\\
\(u_{5}\) \> 8603 \>Nikko Cordial Group  \> Finance \\
 \>  \>  \>  \\
\(u_{6}\) \> 8834 \>Douwa Real Estate  \>Real Estate  \\
\(u_{6}\) \>9501  \>Tokyo Electric Power  \>Electric Power  \\
\(u_{6}\) \>9503  \>Kansai Electric Power  \>Electric Power  \\
\(u_{6}\) \>9504  \>Chuugoku Electric Power  \>Electric Power  \\
\(u_{6}\) \>9506  \>Tohoku Electric Power  \>Electric Power  \\
\(u_{6}\) \>9509  \>Hokkaido Electric Power  \>Electric Power  \\
\(u_{6}\) \> 1804 \>Sato Corporation  \> Construction  \\
\(u_{6}\) \> 1805 \>Tobishima Construction  \> Construction  \\
\(u_{6}\) \> 1806 \>Fujita  \> Construction  \\
\(u_{6}\) \> 1886 \>Aoki Corporation  \> Construction  \\
\>  \>  \>  \\
\(u_{7}\) \>9504  \>Chuugoku Electric Power  \>Electric Power  \\
\(u_{7}\) \>9506  \>Tohoku Electric Power  \>Electric Power  \\
\(u_{7}\) \> 5801 \>Furukawa Electric  \>Nonferrous Metal \\
\(u_{7}\) \> 8004 \>Nichimen  \>Wholesale  \\
\>  \>  \>  \\
\(u_{8}\) \> 8335 \>Ashikaga Bank  \>Bank  \\
\(u_{8}\) \> 9766  \>Konami  \>Service  \\
\>  \>  \>  \\
\(u_{9}\) \> 8004 \>Nichimen  \>Wholesale  \\
\(u_{9}\) \> 8335 \>Ashikaga Bank  \>Bank  \\
\(u_{9}\) \>8752  \>Sumitomo Mitsui Kaijyo   \>Insurance  \\
\end{tabbing}

{\bf TABLE 2} ~~S\&P issues with $Z_i \geq \alpha_{\rm th}$
\begin{tabbing}
xxxxxxxxxxxxx \= xxxxxxxxxx \= xxxxxxxxxxxxxxxxxxxxxxxxxxxxxx \=xxxxxxxxxxxxxxx \kill \\
{\it Eigenvector} \> {\it Ticker} \> {\it Company Name} \> {\it Industries}  \\
\>\>\> \\
\(u_2\) \> AEP \> American Electric Power \> Electric Power  \\
\(u_2\) \> DUK \> Duke Energy Corporation \> Electric Power, Natural Gas \\
  \>\>\> \\
\(u_3\) \> APC  \> Anadarko Petroleum Corp. \> Oil, Gas\\
\(u_3\) \> BHI  \> Baker Hughes Inc. \> Oil Related\\
\(u_3\) \> XOM  \> Exxon Mobil Corporation \> Oil, Coal, Copper\\
\(u_3\) \> HAL  \> Halliburton Company \> Oil, Gas \\
\(u_3\) \> RD  \> Royal Dutch Petroleum Co. \> Oil, Gas, Chemical\\
\(u_3\) \> SLB  \> Schlumberger Ltd. \> Oil \\
\(u_3\) \>  UCL    \> Unocal Corporation     \>  Oil, Gas    \\ 
\>      \>      \>    \\
\(u_4\) \>  GP    \> Georgia-Pacific Group     \>  Paper Manufacture, Pulp        \\ 
\(u_4\) \>  IP    \>  International Paper Co.    \>   Paper Manufacture         \\ 
\(u_4\) \> MEA   \> Mead Corporation     \>  Paper Manufacture, Pulp, Gum   \\ 
\(u_4\) \>  WY \> Weyerhaeuser Company \> Paper Manufacture , Pulp, Forestry, Wooden Goods\\
  \>      \>      \>         \\     
\(u_5\) \>  MRK    \>  Merck \& Co.,  Inc.     \>    Medicine Manufacture        \\ 
\(u_5\) \>   PFE   \>  Pfizer Inc.     \>     Medicine Manufacture       \\ 
\(u_5\) \>  SGP    \>    Schering-Plough Corp.  \>  Medicine Manufacture       \\ 
 \>      \>      \>            \\ 
\(u_6\) \>  BK    \> Bank of New York Co.     \>   Bank     \\ 
\(u_6\) \>  JPM     \>   J.P. Morgan Chase \& Co.    \>   Finance        \\ 
\(u_6\) \>   PNC    \>    PNC Financial Services    \>    Finance        \\ 
\(u_6\) \>   STI    \>  SunTrust Banks,  Inc.    \>  Bank         \\ 
 \>      \>      \>            \\ 
\(u_7\) \> ABX     \>  Barrick Gold Corp.    \>     Gold Mining,  Gold Goods      \\ 
\(u_7\) \> HM      \> Homestake Mining Co.     \>    Gold Mining       \\ 
\(u_7\) \> NEM     \>  Newmont Mining Corp.    \>    Gold Mining        \\ 
\(u_7\) \>   PDG     \>  Placer Dome Inc.    \>     Gold Mining       \\ 
\>      \>      \>            \\ 
\(u_8\) \> SBC  \> SBC Communications Inc. \>  Telecommunication,  Cable Television, Internet  \\ 
\(u_8\) \> VZ     \> Verizon Communications \>     Telecommunication, Internet \\ 
\(u_8\) \> MU   \>  Micron Technology,  Inc.    \>   Semiconductor     \\ 
\(u_8\) \>  TXN \> Texas Instruments \> Semiconductor     \\ 
 \>      \>      \>            \\ 
\(u_9\) \>  AMR    \> AMR Corporation      \> Aviation          \\ 
\(u_9\) \>  DAL    \>  Delta Air Lines,  Inc.    \>  Aviation         \\ 
\(u_9\) \> F \>    Ford Motor Company \>  Automobile\\
\(u_9\) \> GM \> General Motors Corp.  \> Automobile  \\
 \>  \>  \>   \\
\(u_{10}\) \>EIX  \>Edison International  \> Holding Company of Electric Power \\
\(u_{10}\) \>PCG   \>PG\&E Corporation      \> Holding Company of Electric Power  \\
\>  \>  \>   \\
\(u_{11}\) \>AL  \>Alcan Inc.   \> Aluminium,  Aluminium Can  \\
\(u_{11}\) \>AA  \> Alcoa,  Inc.  \> Aluminium  \\
\>  \>  \>   \\
\>  \>  \>   \\
\>  \>  \>   \\
\end{tabbing}

In S\&P,  Electric Power sector and,  Oil and Gas related  sectors 
play major parts in the correlations. In TSE,  Electric Products sector 
and Construction sector play major parts. 

In S\&P,  each eigenvector corresponds to an industrial sector.
This means that 
each industrial sector forms a strongly correlated group. 
On the other hand, in TSE,  there are eigenvectors whose participants 
are from different industrial sectors,  which may indicate a more complicated 
correlation structure of the market.
Thus it seems that TSE and S\&P (NYSE)  have some  
differences in the structure of the correlations, 
while the ``random'' part is well described by the universal theory
in the both markets.
It would be interesting to find the origin of the
difference. This might be useful to give some insights into the difference 
of the economic structures of the two countries.  

As far as our data samples are concerned,  
we may conclude that the method which we propose 
utilizing  small eigenvectors 
and their IPR effectively 
distinguishes strongly correlated groups in the markets. 

We noticed that Giada \etal investigated the grouping of S\&P data in 
Ref.~\cite{giada-marsili} based on a 
model considered by Noh~\cite{noh}. 
The method proposed in Refs.~\cite{giada-marsili,noh} has the 
advantage of directly giving the ``noise-undressed'' correlation 
matrix.  However, the basic assumption of their method is that 
each issue belongs to only one cluster of correlated issues. 
This assumption is apparently not quite true according to 
our analysis. For example, ``Tohoku Electric Power'' appears 
in three different groups in {\bf TABLE 1}. 
Therefore we believe that more analysis based on conservative 
assumptions should be made before applying the estimated 
``true'' correlation to the portfolio management.

\section{Conclusions}
We analyzed the eigenvalues and the eigenvectors of the cross  
correlation matrices of TSE and NYSE (S\&P500) stock market data. 
We found that  results of Refs.~\cite{laloux,plerou,plerou2} reported for 
NYSE are also valid for  TSE. 
The eigenvalue distribution obeys the RMT prediction in the bulk but 
there are some deviations at the large eigenvalues.
We also examined the nearest neighbor  spacing, 
the next-nearest neighbor spacing and the rigidity of the 
eigenvalues  and found that they follow the 
universality of GOE.
These  are consistent with Refs.~\cite{laloux,plerou,plerou2} and 
imply that the large eigenvalues are due to the existence of 
correlations while the eigenvalues distributed in the bulk are 
due to randomness. 
We also examined IPR of the eigenvectors of the correlation matrices.
In the bulk,  IPR distribution follows the prediction  of GOE,  
but there are deviations outside the RMT bounds.  
Plerou \etal~\cite{plerou,plerou2} 
argued that deviations at the lower edge are 
due to strong correlations.
We found that this reasoning is qualitatively valid,  
but quantitatively it cannot explain the fact that 
small eigenvalues with large IPR concentrate 
at the lower edge and 
the observed eigenvalues are smaller than the expected values.

To explain this phenomenon,  we studied RMT with 
deterministic correlations.  We found that each eigenvalue 
from deterministic correlations is observed at 
values which are repelled from the bulk distribution. 
We interpreted this repulsion as a reminiscent of the effect of 
randomness,  known as ``level repulsion''. 
This effect is shown to be deduced 
by solving  the Dyson-type equation numerically.

We also proposed a method to distinguish strongly correlated 
groups of issues based on IPR. It reduces 
the accidental appearance of uncorrelated issues. 
Applying this method,  we found that issues of S\&P  are grouped 
according to the industrial sectors. On the other hand,  
issues of TSE are grouped in more complicated ways,  suggesting 
 some differences in the structure of the markets.
\\

{\it Acknowledgement}  The authors acknowledge Institute for 
Asset Management of Mizuho T.~B. and Hiraku Kusaka at BNP Paribas for 
providing the stock price data. They thank Hiraku Kusaka also for 
his critical reading of the manuscript and useful comments, 
and Shinobu Hikami for discussions. K.I. has benefited from the Grand-in-Aid 
for Science(B),No.1430114 of JSPS.\\

\end{document}

%% file: graph10-1n.tex
\setlength{\unitlength}{0.240900pt}
\ifx\plotpoint\undefined\newsavebox{\plotpoint}\fi
\begin{picture}(1500,900)(0,0)
\font\gnuplot=cmr10 at 10pt
\gnuplot
\sbox{\plotpoint}{\rule[-0.200pt]{0.400pt}{0.400pt}}%
\put(201.0,123.0){\rule[-0.200pt]{4.818pt}{0.400pt}}
\put(181,123){\makebox(0,0)[r]{ 0}}
\put(1419.0,123.0){\rule[-0.200pt]{4.818pt}{0.400pt}}
\put(201.0,205.0){\rule[-0.200pt]{4.818pt}{0.400pt}}
\put(181,205){\makebox(0,0)[r]{ 0.05}}
\put(1419.0,205.0){\rule[-0.200pt]{4.818pt}{0.400pt}}
\put(201.0,287.0){\rule[-0.200pt]{4.818pt}{0.400pt}}
\put(181,287){\makebox(0,0)[r]{ 0.1}}
\put(1419.0,287.0){\rule[-0.200pt]{4.818pt}{0.400pt}}
\put(201.0,369.0){\rule[-0.200pt]{4.818pt}{0.400pt}}
\put(181,369){\makebox(0,0)[r]{ 0.15}}
\put(1419.0,369.0){\rule[-0.200pt]{4.818pt}{0.400pt}}
\put(201.0,451.0){\rule[-0.200pt]{4.818pt}{0.400pt}}
\put(181,451){\makebox(0,0)[r]{ 0.2}}
\put(1419.0,451.0){\rule[-0.200pt]{4.818pt}{0.400pt}}
\put(201.0,532.0){\rule[-0.200pt]{4.818pt}{0.400pt}}
\put(181,532){\makebox(0,0)[r]{ 0.25}}
\put(1419.0,532.0){\rule[-0.200pt]{4.818pt}{0.400pt}}
\put(201.0,614.0){\rule[-0.200pt]{4.818pt}{0.400pt}}
\put(181,614){\makebox(0,0)[r]{ 0.3}}
\put(1419.0,614.0){\rule[-0.200pt]{4.818pt}{0.400pt}}
\put(201.0,696.0){\rule[-0.200pt]{4.818pt}{0.400pt}}
\put(181,696){\makebox(0,0)[r]{ 0.35}}
\put(1419.0,696.0){\rule[-0.200pt]{4.818pt}{0.400pt}}
\put(201.0,778.0){\rule[-0.200pt]{4.818pt}{0.400pt}}
\put(181,778){\makebox(0,0)[r]{ 0.4}}
\put(1419.0,778.0){\rule[-0.200pt]{4.818pt}{0.400pt}}
\put(201.0,860.0){\rule[-0.200pt]{4.818pt}{0.400pt}}
\put(181,860){\makebox(0,0)[r]{ 0.45}}
\put(1419.0,860.0){\rule[-0.200pt]{4.818pt}{0.400pt}}
\put(201.0,123.0){\rule[-0.200pt]{0.400pt}{4.818pt}}
\put(201,82){\makebox(0,0){ 0.1}}
\put(201.0,840.0){\rule[-0.200pt]{0.400pt}{4.818pt}}
\put(407.0,123.0){\rule[-0.200pt]{0.400pt}{4.818pt}}
\put(407,82){\makebox(0,0){ 0.15}}
\put(407.0,840.0){\rule[-0.200pt]{0.400pt}{4.818pt}}
\put(614.0,123.0){\rule[-0.200pt]{0.400pt}{4.818pt}}
\put(614,82){\makebox(0,0){ 0.2}}
\put(614.0,840.0){\rule[-0.200pt]{0.400pt}{4.818pt}}
\put(820.0,123.0){\rule[-0.200pt]{0.400pt}{4.818pt}}
\put(820,82){\makebox(0,0){ 0.25}}
\put(820.0,840.0){\rule[-0.200pt]{0.400pt}{4.818pt}}
\put(1026.0,123.0){\rule[-0.200pt]{0.400pt}{4.818pt}}
\put(1026,82){\makebox(0,0){ 0.3}}
\put(1026.0,840.0){\rule[-0.200pt]{0.400pt}{4.818pt}}
\put(1233.0,123.0){\rule[-0.200pt]{0.400pt}{4.818pt}}
\put(1233,82){\makebox(0,0){ 0.35}}
\put(1233.0,840.0){\rule[-0.200pt]{0.400pt}{4.818pt}}
\put(1439.0,123.0){\rule[-0.200pt]{0.400pt}{4.818pt}}
\put(1439,82){\makebox(0,0){ 0.4}}
\put(1439.0,840.0){\rule[-0.200pt]{0.400pt}{4.818pt}}
\put(201.0,123.0){\rule[-0.200pt]{298.234pt}{0.400pt}}
\put(1439.0,123.0){\rule[-0.200pt]{0.400pt}{177.543pt}}
\put(201.0,860.0){\rule[-0.200pt]{298.234pt}{0.400pt}}
\put(40,491){\makebox(0,0){${\LARGE C}$}}
\put(820,21){\makebox(0,0){${\LARGE \widetilde{C}}$}}
\put(201.0,123.0){\rule[-0.200pt]{0.400pt}{177.543pt}}
\put(201,169){\makebox(0,0){$+$}}
\put(242,175){\makebox(0,0){$+$}}
\put(284,180){\makebox(0,0){$+$}}
\put(325,187){\makebox(0,0){$+$}}
\put(366,193){\makebox(0,0){$+$}}
\put(407,198){\makebox(0,0){$+$}}
\put(449,205){\makebox(0,0){$+$}}
\put(490,211){\makebox(0,0){$+$}}
\put(531,216){\makebox(0,0){$+$}}
\put(572,223){\makebox(0,0){$+$}}
\put(614,229){\makebox(0,0){$+$}}
\put(655,238){\makebox(0,0){$+$}}
\put(696,244){\makebox(0,0){$+$}}
\put(737,251){\makebox(0,0){$+$}}
\put(779,251){\makebox(0,0){$+$}}
\put(820,259){\makebox(0,0){$+$}}
\put(861,265){\makebox(0,0){$+$}}
\put(903,275){\makebox(0,0){$+$}}
\put(944,282){\makebox(0,0){$+$}}
\put(985,292){\makebox(0,0){$+$}}
\put(1026,302){\makebox(0,0){$+$}}
\put(1068,310){\makebox(0,0){$+$}}
\put(1109,320){\makebox(0,0){$+$}}
\put(1150,329){\makebox(0,0){$+$}}
\put(1191,339){\makebox(0,0){$+$}}
\put(1233,349){\makebox(0,0){$+$}}
\put(1274,362){\makebox(0,0){$+$}}
\put(201,287){\usebox{\plotpoint}}
\multiput(201.00,287.58)(1.298,0.494){29}{\rule{1.125pt}{0.119pt}}
\multiput(201.00,286.17)(38.665,16.000){2}{\rule{0.563pt}{0.400pt}}
\multiput(242.00,303.58)(1.249,0.495){31}{\rule{1.088pt}{0.119pt}}
\multiput(242.00,302.17)(39.741,17.000){2}{\rule{0.544pt}{0.400pt}}
\multiput(284.00,320.58)(1.298,0.494){29}{\rule{1.125pt}{0.119pt}}
\multiput(284.00,319.17)(38.665,16.000){2}{\rule{0.563pt}{0.400pt}}
\multiput(325.00,336.58)(1.298,0.494){29}{\rule{1.125pt}{0.119pt}}
\multiput(325.00,335.17)(38.665,16.000){2}{\rule{0.563pt}{0.400pt}}
\multiput(366.00,352.58)(1.219,0.495){31}{\rule{1.065pt}{0.119pt}}
\multiput(366.00,351.17)(38.790,17.000){2}{\rule{0.532pt}{0.400pt}}
\multiput(407.00,369.58)(1.330,0.494){29}{\rule{1.150pt}{0.119pt}}
\multiput(407.00,368.17)(39.613,16.000){2}{\rule{0.575pt}{0.400pt}}
\multiput(449.00,385.58)(1.298,0.494){29}{\rule{1.125pt}{0.119pt}}
\multiput(449.00,384.17)(38.665,16.000){2}{\rule{0.563pt}{0.400pt}}
\multiput(490.00,401.58)(1.219,0.495){31}{\rule{1.065pt}{0.119pt}}
\multiput(490.00,400.17)(38.790,17.000){2}{\rule{0.532pt}{0.400pt}}
\multiput(531.00,418.58)(1.298,0.494){29}{\rule{1.125pt}{0.119pt}}
\multiput(531.00,417.17)(38.665,16.000){2}{\rule{0.563pt}{0.400pt}}
\multiput(572.00,434.58)(1.249,0.495){31}{\rule{1.088pt}{0.119pt}}
\multiput(572.00,433.17)(39.741,17.000){2}{\rule{0.544pt}{0.400pt}}
\multiput(614.00,451.58)(1.298,0.494){29}{\rule{1.125pt}{0.119pt}}
\multiput(614.00,450.17)(38.665,16.000){2}{\rule{0.563pt}{0.400pt}}
\multiput(655.00,467.58)(1.298,0.494){29}{\rule{1.125pt}{0.119pt}}
\multiput(655.00,466.17)(38.665,16.000){2}{\rule{0.563pt}{0.400pt}}
\multiput(696.00,483.58)(1.219,0.495){31}{\rule{1.065pt}{0.119pt}}
\multiput(696.00,482.17)(38.790,17.000){2}{\rule{0.532pt}{0.400pt}}
\multiput(737.00,500.58)(1.330,0.494){29}{\rule{1.150pt}{0.119pt}}
\multiput(737.00,499.17)(39.613,16.000){2}{\rule{0.575pt}{0.400pt}}
\multiput(779.00,516.58)(1.298,0.494){29}{\rule{1.125pt}{0.119pt}}
\multiput(779.00,515.17)(38.665,16.000){2}{\rule{0.563pt}{0.400pt}}
\multiput(820.00,532.58)(1.219,0.495){31}{\rule{1.065pt}{0.119pt}}
\multiput(820.00,531.17)(38.790,17.000){2}{\rule{0.532pt}{0.400pt}}
\multiput(861.00,549.58)(1.330,0.494){29}{\rule{1.150pt}{0.119pt}}
\multiput(861.00,548.17)(39.613,16.000){2}{\rule{0.575pt}{0.400pt}}
\multiput(903.00,565.58)(1.219,0.495){31}{\rule{1.065pt}{0.119pt}}
\multiput(903.00,564.17)(38.790,17.000){2}{\rule{0.532pt}{0.400pt}}
\multiput(944.00,582.58)(1.298,0.494){29}{\rule{1.125pt}{0.119pt}}
\multiput(944.00,581.17)(38.665,16.000){2}{\rule{0.563pt}{0.400pt}}
\multiput(985.00,598.58)(1.298,0.494){29}{\rule{1.125pt}{0.119pt}}
\multiput(985.00,597.17)(38.665,16.000){2}{\rule{0.563pt}{0.400pt}}
\multiput(1026.00,614.58)(1.249,0.495){31}{\rule{1.088pt}{0.119pt}}
\multiput(1026.00,613.17)(39.741,17.000){2}{\rule{0.544pt}{0.400pt}}
\multiput(1068.00,631.58)(1.298,0.494){29}{\rule{1.125pt}{0.119pt}}
\multiput(1068.00,630.17)(38.665,16.000){2}{\rule{0.563pt}{0.400pt}}
\multiput(1109.00,647.58)(1.298,0.494){29}{\rule{1.125pt}{0.119pt}}
\multiput(1109.00,646.17)(38.665,16.000){2}{\rule{0.563pt}{0.400pt}}
\multiput(1150.00,663.58)(1.219,0.495){31}{\rule{1.065pt}{0.119pt}}
\multiput(1150.00,662.17)(38.790,17.000){2}{\rule{0.532pt}{0.400pt}}
\multiput(1191.00,680.58)(1.330,0.494){29}{\rule{1.150pt}{0.119pt}}
\multiput(1191.00,679.17)(39.613,16.000){2}{\rule{0.575pt}{0.400pt}}
\multiput(1233.00,696.58)(1.219,0.495){31}{\rule{1.065pt}{0.119pt}}
\multiput(1233.00,695.17)(38.790,17.000){2}{\rule{0.532pt}{0.400pt}}
\multiput(1274.00,713.58)(1.298,0.494){29}{\rule{1.125pt}{0.119pt}}
\multiput(1274.00,712.17)(38.665,16.000){2}{\rule{0.563pt}{0.400pt}}
\multiput(1315.00,729.58)(1.298,0.494){29}{\rule{1.125pt}{0.119pt}}
\multiput(1315.00,728.17)(38.665,16.000){2}{\rule{0.563pt}{0.400pt}}
\multiput(1356.00,745.58)(1.249,0.495){31}{\rule{1.088pt}{0.119pt}}
\multiput(1356.00,744.17)(39.741,17.000){2}{\rule{0.544pt}{0.400pt}}
\multiput(1398.00,762.58)(1.298,0.494){29}{\rule{1.125pt}{0.119pt}}
\multiput(1398.00,761.17)(38.665,16.000){2}{\rule{0.563pt}{0.400pt}}
\put(1439,778){\usebox{\plotpoint}}
\end{picture}

%% file: graph10-2n.tex
\setlength{\unitlength}{0.240900pt}
\ifx\plotpoint\undefined\newsavebox{\plotpoint}\fi
\begin{picture}(1500,900)(0,0)
\font\gnuplot=cmr10 at 10pt
\gnuplot
\sbox{\plotpoint}{\rule[-0.200pt]{0.400pt}{0.400pt}}%
\put(181.0,123.0){\rule[-0.200pt]{4.818pt}{0.400pt}}
\put(161,123){\makebox(0,0)[r]{ 2.5}}
\put(1419.0,123.0){\rule[-0.200pt]{4.818pt}{0.400pt}}
\put(181.0,270.0){\rule[-0.200pt]{4.818pt}{0.400pt}}
\put(161,270){\makebox(0,0)[r]{ 3}}
\put(1419.0,270.0){\rule[-0.200pt]{4.818pt}{0.400pt}}
\put(181.0,418.0){\rule[-0.200pt]{4.818pt}{0.400pt}}
\put(161,418){\makebox(0,0)[r]{ 3.5}}
\put(1419.0,418.0){\rule[-0.200pt]{4.818pt}{0.400pt}}
\put(181.0,565.0){\rule[-0.200pt]{4.818pt}{0.400pt}}
\put(161,565){\makebox(0,0)[r]{ 4}}
\put(1419.0,565.0){\rule[-0.200pt]{4.818pt}{0.400pt}}
\put(181.0,713.0){\rule[-0.200pt]{4.818pt}{0.400pt}}
\put(161,713){\makebox(0,0)[r]{ 4.5}}
\put(1419.0,713.0){\rule[-0.200pt]{4.818pt}{0.400pt}}
\put(181.0,860.0){\rule[-0.200pt]{4.818pt}{0.400pt}}
\put(161,860){\makebox(0,0)[r]{ 5}}
\put(1419.0,860.0){\rule[-0.200pt]{4.818pt}{0.400pt}}
\put(181.0,123.0){\rule[-0.200pt]{0.400pt}{4.818pt}}
\put(181,82){\makebox(0,0){ 2.5}}
\put(181.0,840.0){\rule[-0.200pt]{0.400pt}{4.818pt}}
\put(496.0,123.0){\rule[-0.200pt]{0.400pt}{4.818pt}}
\put(496,82){\makebox(0,0){ 3}}
\put(496.0,840.0){\rule[-0.200pt]{0.400pt}{4.818pt}}
\put(810.0,123.0){\rule[-0.200pt]{0.400pt}{4.818pt}}
\put(810,82){\makebox(0,0){ 3.5}}
\put(810.0,840.0){\rule[-0.200pt]{0.400pt}{4.818pt}}
\put(1125.0,123.0){\rule[-0.200pt]{0.400pt}{4.818pt}}
\put(1125,82){\makebox(0,0){ 4}}
\put(1125.0,840.0){\rule[-0.200pt]{0.400pt}{4.818pt}}
\put(1439.0,123.0){\rule[-0.200pt]{0.400pt}{4.818pt}}
\put(1439,82){\makebox(0,0){ 4.5}}
\put(1439.0,840.0){\rule[-0.200pt]{0.400pt}{4.818pt}}
\put(181.0,123.0){\rule[-0.200pt]{303.052pt}{0.400pt}}
\put(1439.0,123.0){\rule[-0.200pt]{0.400pt}{177.543pt}}
\put(181.0,860.0){\rule[-0.200pt]{303.052pt}{0.400pt}}
\put(40,491){\makebox(0,0){${\LARGE C}$}}
\put(810,21){\makebox(0,0){${\LARGE \widetilde{C}}$}}
\put(181.0,123.0){\rule[-0.200pt]{0.400pt}{177.543pt}}
\put(204,292){\makebox(0,0){$+$}}
\put(301,324){\makebox(0,0){$+$}}
\put(407,365){\makebox(0,0){$+$}}
\put(518,410){\makebox(0,0){$+$}}
\put(632,457){\makebox(0,0){$+$}}
\put(749,508){\makebox(0,0){$+$}}
\put(867,560){\makebox(0,0){$+$}}
\put(986,612){\makebox(0,0){$+$}}
\put(1107,667){\makebox(0,0){$+$}}
\put(1228,721){\makebox(0,0){$+$}}
\multiput(181.00,123.61)(1.132,0.447){3}{\rule{0.900pt}{0.108pt}}
\multiput(181.00,122.17)(4.132,3.000){2}{\rule{0.450pt}{0.400pt}}
\multiput(187.00,126.61)(1.355,0.447){3}{\rule{1.033pt}{0.108pt}}
\multiput(187.00,125.17)(4.855,3.000){2}{\rule{0.517pt}{0.400pt}}
\multiput(194.00,129.61)(1.132,0.447){3}{\rule{0.900pt}{0.108pt}}
\multiput(194.00,128.17)(4.132,3.000){2}{\rule{0.450pt}{0.400pt}}
\multiput(200.00,132.61)(1.132,0.447){3}{\rule{0.900pt}{0.108pt}}
\multiput(200.00,131.17)(4.132,3.000){2}{\rule{0.450pt}{0.400pt}}
\multiput(206.00,135.61)(1.132,0.447){3}{\rule{0.900pt}{0.108pt}}
\multiput(206.00,134.17)(4.132,3.000){2}{\rule{0.450pt}{0.400pt}}
\multiput(212.00,138.61)(1.355,0.447){3}{\rule{1.033pt}{0.108pt}}
\multiput(212.00,137.17)(4.855,3.000){2}{\rule{0.517pt}{0.400pt}}
\multiput(219.00,141.61)(1.132,0.447){3}{\rule{0.900pt}{0.108pt}}
\multiput(219.00,140.17)(4.132,3.000){2}{\rule{0.450pt}{0.400pt}}
\multiput(225.00,144.61)(1.132,0.447){3}{\rule{0.900pt}{0.108pt}}
\multiput(225.00,143.17)(4.132,3.000){2}{\rule{0.450pt}{0.400pt}}
\multiput(231.00,147.61)(1.355,0.447){3}{\rule{1.033pt}{0.108pt}}
\multiput(231.00,146.17)(4.855,3.000){2}{\rule{0.517pt}{0.400pt}}
\put(238,150.17){\rule{1.300pt}{0.400pt}}
\multiput(238.00,149.17)(3.302,2.000){2}{\rule{0.650pt}{0.400pt}}
\multiput(244.00,152.61)(1.132,0.447){3}{\rule{0.900pt}{0.108pt}}
\multiput(244.00,151.17)(4.132,3.000){2}{\rule{0.450pt}{0.400pt}}
\multiput(250.00,155.61)(1.132,0.447){3}{\rule{0.900pt}{0.108pt}}
\multiput(250.00,154.17)(4.132,3.000){2}{\rule{0.450pt}{0.400pt}}
\multiput(256.00,158.61)(1.355,0.447){3}{\rule{1.033pt}{0.108pt}}
\multiput(256.00,157.17)(4.855,3.000){2}{\rule{0.517pt}{0.400pt}}
\multiput(263.00,161.61)(1.132,0.447){3}{\rule{0.900pt}{0.108pt}}
\multiput(263.00,160.17)(4.132,3.000){2}{\rule{0.450pt}{0.400pt}}
\multiput(269.00,164.61)(1.132,0.447){3}{\rule{0.900pt}{0.108pt}}
\multiput(269.00,163.17)(4.132,3.000){2}{\rule{0.450pt}{0.400pt}}
\multiput(275.00,167.61)(1.355,0.447){3}{\rule{1.033pt}{0.108pt}}
\multiput(275.00,166.17)(4.855,3.000){2}{\rule{0.517pt}{0.400pt}}
\multiput(282.00,170.61)(1.132,0.447){3}{\rule{0.900pt}{0.108pt}}
\multiput(282.00,169.17)(4.132,3.000){2}{\rule{0.450pt}{0.400pt}}
\multiput(288.00,173.61)(1.132,0.447){3}{\rule{0.900pt}{0.108pt}}
\multiput(288.00,172.17)(4.132,3.000){2}{\rule{0.450pt}{0.400pt}}
\multiput(294.00,176.61)(1.355,0.447){3}{\rule{1.033pt}{0.108pt}}
\multiput(294.00,175.17)(4.855,3.000){2}{\rule{0.517pt}{0.400pt}}
\multiput(301.00,179.61)(1.132,0.447){3}{\rule{0.900pt}{0.108pt}}
\multiput(301.00,178.17)(4.132,3.000){2}{\rule{0.450pt}{0.400pt}}
\multiput(307.00,182.61)(1.132,0.447){3}{\rule{0.900pt}{0.108pt}}
\multiput(307.00,181.17)(4.132,3.000){2}{\rule{0.450pt}{0.400pt}}
\multiput(313.00,185.61)(1.132,0.447){3}{\rule{0.900pt}{0.108pt}}
\multiput(313.00,184.17)(4.132,3.000){2}{\rule{0.450pt}{0.400pt}}
\multiput(319.00,188.61)(1.355,0.447){3}{\rule{1.033pt}{0.108pt}}
\multiput(319.00,187.17)(4.855,3.000){2}{\rule{0.517pt}{0.400pt}}
\multiput(326.00,191.61)(1.132,0.447){3}{\rule{0.900pt}{0.108pt}}
\multiput(326.00,190.17)(4.132,3.000){2}{\rule{0.450pt}{0.400pt}}
\multiput(332.00,194.61)(1.132,0.447){3}{\rule{0.900pt}{0.108pt}}
\multiput(332.00,193.17)(4.132,3.000){2}{\rule{0.450pt}{0.400pt}}
\multiput(338.00,197.61)(1.355,0.447){3}{\rule{1.033pt}{0.108pt}}
\multiput(338.00,196.17)(4.855,3.000){2}{\rule{0.517pt}{0.400pt}}
\multiput(345.00,200.61)(1.132,0.447){3}{\rule{0.900pt}{0.108pt}}
\multiput(345.00,199.17)(4.132,3.000){2}{\rule{0.450pt}{0.400pt}}
\multiput(351.00,203.61)(1.132,0.447){3}{\rule{0.900pt}{0.108pt}}
\multiput(351.00,202.17)(4.132,3.000){2}{\rule{0.450pt}{0.400pt}}
\put(357,206.17){\rule{1.300pt}{0.400pt}}
\multiput(357.00,205.17)(3.302,2.000){2}{\rule{0.650pt}{0.400pt}}
\multiput(363.00,208.61)(1.355,0.447){3}{\rule{1.033pt}{0.108pt}}
\multiput(363.00,207.17)(4.855,3.000){2}{\rule{0.517pt}{0.400pt}}
\multiput(370.00,211.61)(1.132,0.447){3}{\rule{0.900pt}{0.108pt}}
\multiput(370.00,210.17)(4.132,3.000){2}{\rule{0.450pt}{0.400pt}}
\multiput(376.00,214.61)(1.132,0.447){3}{\rule{0.900pt}{0.108pt}}
\multiput(376.00,213.17)(4.132,3.000){2}{\rule{0.450pt}{0.400pt}}
\multiput(382.00,217.61)(1.355,0.447){3}{\rule{1.033pt}{0.108pt}}
\multiput(382.00,216.17)(4.855,3.000){2}{\rule{0.517pt}{0.400pt}}
\multiput(389.00,220.61)(1.132,0.447){3}{\rule{0.900pt}{0.108pt}}
\multiput(389.00,219.17)(4.132,3.000){2}{\rule{0.450pt}{0.400pt}}
\multiput(395.00,223.61)(1.132,0.447){3}{\rule{0.900pt}{0.108pt}}
\multiput(395.00,222.17)(4.132,3.000){2}{\rule{0.450pt}{0.400pt}}
\multiput(401.00,226.61)(1.132,0.447){3}{\rule{0.900pt}{0.108pt}}
\multiput(401.00,225.17)(4.132,3.000){2}{\rule{0.450pt}{0.400pt}}
\multiput(407.00,229.61)(1.355,0.447){3}{\rule{1.033pt}{0.108pt}}
\multiput(407.00,228.17)(4.855,3.000){2}{\rule{0.517pt}{0.400pt}}
\multiput(414.00,232.61)(1.132,0.447){3}{\rule{0.900pt}{0.108pt}}
\multiput(414.00,231.17)(4.132,3.000){2}{\rule{0.450pt}{0.400pt}}
\multiput(420.00,235.61)(1.132,0.447){3}{\rule{0.900pt}{0.108pt}}
\multiput(420.00,234.17)(4.132,3.000){2}{\rule{0.450pt}{0.400pt}}
\multiput(426.00,238.61)(1.355,0.447){3}{\rule{1.033pt}{0.108pt}}
\multiput(426.00,237.17)(4.855,3.000){2}{\rule{0.517pt}{0.400pt}}
\multiput(433.00,241.61)(1.132,0.447){3}{\rule{0.900pt}{0.108pt}}
\multiput(433.00,240.17)(4.132,3.000){2}{\rule{0.450pt}{0.400pt}}
\multiput(439.00,244.61)(1.132,0.447){3}{\rule{0.900pt}{0.108pt}}
\multiput(439.00,243.17)(4.132,3.000){2}{\rule{0.450pt}{0.400pt}}
\multiput(445.00,247.61)(1.132,0.447){3}{\rule{0.900pt}{0.108pt}}
\multiput(445.00,246.17)(4.132,3.000){2}{\rule{0.450pt}{0.400pt}}
\multiput(451.00,250.61)(1.355,0.447){3}{\rule{1.033pt}{0.108pt}}
\multiput(451.00,249.17)(4.855,3.000){2}{\rule{0.517pt}{0.400pt}}
\multiput(458.00,253.61)(1.132,0.447){3}{\rule{0.900pt}{0.108pt}}
\multiput(458.00,252.17)(4.132,3.000){2}{\rule{0.450pt}{0.400pt}}
\multiput(464.00,256.61)(1.132,0.447){3}{\rule{0.900pt}{0.108pt}}
\multiput(464.00,255.17)(4.132,3.000){2}{\rule{0.450pt}{0.400pt}}
\multiput(470.00,259.61)(1.355,0.447){3}{\rule{1.033pt}{0.108pt}}
\multiput(470.00,258.17)(4.855,3.000){2}{\rule{0.517pt}{0.400pt}}
\multiput(477.00,262.61)(1.132,0.447){3}{\rule{0.900pt}{0.108pt}}
\multiput(477.00,261.17)(4.132,3.000){2}{\rule{0.450pt}{0.400pt}}
\put(483,265.17){\rule{1.300pt}{0.400pt}}
\multiput(483.00,264.17)(3.302,2.000){2}{\rule{0.650pt}{0.400pt}}
\multiput(489.00,267.61)(1.132,0.447){3}{\rule{0.900pt}{0.108pt}}
\multiput(489.00,266.17)(4.132,3.000){2}{\rule{0.450pt}{0.400pt}}
\multiput(495.00,270.61)(1.355,0.447){3}{\rule{1.033pt}{0.108pt}}
\multiput(495.00,269.17)(4.855,3.000){2}{\rule{0.517pt}{0.400pt}}
\multiput(502.00,273.61)(1.132,0.447){3}{\rule{0.900pt}{0.108pt}}
\multiput(502.00,272.17)(4.132,3.000){2}{\rule{0.450pt}{0.400pt}}
\multiput(508.00,276.61)(1.132,0.447){3}{\rule{0.900pt}{0.108pt}}
\multiput(508.00,275.17)(4.132,3.000){2}{\rule{0.450pt}{0.400pt}}
\multiput(514.00,279.61)(1.355,0.447){3}{\rule{1.033pt}{0.108pt}}
\multiput(514.00,278.17)(4.855,3.000){2}{\rule{0.517pt}{0.400pt}}
\multiput(521.00,282.61)(1.132,0.447){3}{\rule{0.900pt}{0.108pt}}
\multiput(521.00,281.17)(4.132,3.000){2}{\rule{0.450pt}{0.400pt}}
\multiput(527.00,285.61)(1.132,0.447){3}{\rule{0.900pt}{0.108pt}}
\multiput(527.00,284.17)(4.132,3.000){2}{\rule{0.450pt}{0.400pt}}
\multiput(533.00,288.61)(1.355,0.447){3}{\rule{1.033pt}{0.108pt}}
\multiput(533.00,287.17)(4.855,3.000){2}{\rule{0.517pt}{0.400pt}}
\multiput(540.00,291.61)(1.132,0.447){3}{\rule{0.900pt}{0.108pt}}
\multiput(540.00,290.17)(4.132,3.000){2}{\rule{0.450pt}{0.400pt}}
\multiput(546.00,294.61)(1.132,0.447){3}{\rule{0.900pt}{0.108pt}}
\multiput(546.00,293.17)(4.132,3.000){2}{\rule{0.450pt}{0.400pt}}
\multiput(552.00,297.61)(1.132,0.447){3}{\rule{0.900pt}{0.108pt}}
\multiput(552.00,296.17)(4.132,3.000){2}{\rule{0.450pt}{0.400pt}}
\multiput(558.00,300.61)(1.355,0.447){3}{\rule{1.033pt}{0.108pt}}
\multiput(558.00,299.17)(4.855,3.000){2}{\rule{0.517pt}{0.400pt}}
\multiput(565.00,303.61)(1.132,0.447){3}{\rule{0.900pt}{0.108pt}}
\multiput(565.00,302.17)(4.132,3.000){2}{\rule{0.450pt}{0.400pt}}
\multiput(571.00,306.61)(1.132,0.447){3}{\rule{0.900pt}{0.108pt}}
\multiput(571.00,305.17)(4.132,3.000){2}{\rule{0.450pt}{0.400pt}}
\multiput(577.00,309.61)(1.355,0.447){3}{\rule{1.033pt}{0.108pt}}
\multiput(577.00,308.17)(4.855,3.000){2}{\rule{0.517pt}{0.400pt}}
\multiput(584.00,312.61)(1.132,0.447){3}{\rule{0.900pt}{0.108pt}}
\multiput(584.00,311.17)(4.132,3.000){2}{\rule{0.450pt}{0.400pt}}
\multiput(590.00,315.61)(1.132,0.447){3}{\rule{0.900pt}{0.108pt}}
\multiput(590.00,314.17)(4.132,3.000){2}{\rule{0.450pt}{0.400pt}}
\multiput(596.00,318.61)(1.132,0.447){3}{\rule{0.900pt}{0.108pt}}
\multiput(596.00,317.17)(4.132,3.000){2}{\rule{0.450pt}{0.400pt}}
\put(602,321.17){\rule{1.500pt}{0.400pt}}
\multiput(602.00,320.17)(3.887,2.000){2}{\rule{0.750pt}{0.400pt}}
\multiput(609.00,323.61)(1.132,0.447){3}{\rule{0.900pt}{0.108pt}}
\multiput(609.00,322.17)(4.132,3.000){2}{\rule{0.450pt}{0.400pt}}
\multiput(615.00,326.61)(1.132,0.447){3}{\rule{0.900pt}{0.108pt}}
\multiput(615.00,325.17)(4.132,3.000){2}{\rule{0.450pt}{0.400pt}}
\multiput(621.00,329.61)(1.355,0.447){3}{\rule{1.033pt}{0.108pt}}
\multiput(621.00,328.17)(4.855,3.000){2}{\rule{0.517pt}{0.400pt}}
\multiput(628.00,332.61)(1.132,0.447){3}{\rule{0.900pt}{0.108pt}}
\multiput(628.00,331.17)(4.132,3.000){2}{\rule{0.450pt}{0.400pt}}
\multiput(634.00,335.61)(1.132,0.447){3}{\rule{0.900pt}{0.108pt}}
\multiput(634.00,334.17)(4.132,3.000){2}{\rule{0.450pt}{0.400pt}}
\multiput(640.00,338.61)(1.132,0.447){3}{\rule{0.900pt}{0.108pt}}
\multiput(640.00,337.17)(4.132,3.000){2}{\rule{0.450pt}{0.400pt}}
\multiput(646.00,341.61)(1.355,0.447){3}{\rule{1.033pt}{0.108pt}}
\multiput(646.00,340.17)(4.855,3.000){2}{\rule{0.517pt}{0.400pt}}
\multiput(653.00,344.61)(1.132,0.447){3}{\rule{0.900pt}{0.108pt}}
\multiput(653.00,343.17)(4.132,3.000){2}{\rule{0.450pt}{0.400pt}}
\multiput(659.00,347.61)(1.132,0.447){3}{\rule{0.900pt}{0.108pt}}
\multiput(659.00,346.17)(4.132,3.000){2}{\rule{0.450pt}{0.400pt}}
\multiput(665.00,350.61)(1.355,0.447){3}{\rule{1.033pt}{0.108pt}}
\multiput(665.00,349.17)(4.855,3.000){2}{\rule{0.517pt}{0.400pt}}
\multiput(672.00,353.61)(1.132,0.447){3}{\rule{0.900pt}{0.108pt}}
\multiput(672.00,352.17)(4.132,3.000){2}{\rule{0.450pt}{0.400pt}}
\multiput(678.00,356.61)(1.132,0.447){3}{\rule{0.900pt}{0.108pt}}
\multiput(678.00,355.17)(4.132,3.000){2}{\rule{0.450pt}{0.400pt}}
\multiput(684.00,359.61)(1.132,0.447){3}{\rule{0.900pt}{0.108pt}}
\multiput(684.00,358.17)(4.132,3.000){2}{\rule{0.450pt}{0.400pt}}
\multiput(690.00,362.61)(1.355,0.447){3}{\rule{1.033pt}{0.108pt}}
\multiput(690.00,361.17)(4.855,3.000){2}{\rule{0.517pt}{0.400pt}}
\multiput(697.00,365.61)(1.132,0.447){3}{\rule{0.900pt}{0.108pt}}
\multiput(697.00,364.17)(4.132,3.000){2}{\rule{0.450pt}{0.400pt}}
\multiput(703.00,368.61)(1.132,0.447){3}{\rule{0.900pt}{0.108pt}}
\multiput(703.00,367.17)(4.132,3.000){2}{\rule{0.450pt}{0.400pt}}
\multiput(709.00,371.61)(1.355,0.447){3}{\rule{1.033pt}{0.108pt}}
\multiput(709.00,370.17)(4.855,3.000){2}{\rule{0.517pt}{0.400pt}}
\multiput(716.00,374.61)(1.132,0.447){3}{\rule{0.900pt}{0.108pt}}
\multiput(716.00,373.17)(4.132,3.000){2}{\rule{0.450pt}{0.400pt}}
\put(722,377.17){\rule{1.300pt}{0.400pt}}
\multiput(722.00,376.17)(3.302,2.000){2}{\rule{0.650pt}{0.400pt}}
\multiput(728.00,379.61)(1.355,0.447){3}{\rule{1.033pt}{0.108pt}}
\multiput(728.00,378.17)(4.855,3.000){2}{\rule{0.517pt}{0.400pt}}
\multiput(735.00,382.61)(1.132,0.447){3}{\rule{0.900pt}{0.108pt}}
\multiput(735.00,381.17)(4.132,3.000){2}{\rule{0.450pt}{0.400pt}}
\multiput(741.00,385.61)(1.132,0.447){3}{\rule{0.900pt}{0.108pt}}
\multiput(741.00,384.17)(4.132,3.000){2}{\rule{0.450pt}{0.400pt}}
\multiput(747.00,388.61)(1.132,0.447){3}{\rule{0.900pt}{0.108pt}}
\multiput(747.00,387.17)(4.132,3.000){2}{\rule{0.450pt}{0.400pt}}
\multiput(753.00,391.61)(1.355,0.447){3}{\rule{1.033pt}{0.108pt}}
\multiput(753.00,390.17)(4.855,3.000){2}{\rule{0.517pt}{0.400pt}}
\multiput(760.00,394.61)(1.132,0.447){3}{\rule{0.900pt}{0.108pt}}
\multiput(760.00,393.17)(4.132,3.000){2}{\rule{0.450pt}{0.400pt}}
\multiput(766.00,397.61)(1.132,0.447){3}{\rule{0.900pt}{0.108pt}}
\multiput(766.00,396.17)(4.132,3.000){2}{\rule{0.450pt}{0.400pt}}
\multiput(772.00,400.61)(1.355,0.447){3}{\rule{1.033pt}{0.108pt}}
\multiput(772.00,399.17)(4.855,3.000){2}{\rule{0.517pt}{0.400pt}}
\multiput(779.00,403.61)(1.132,0.447){3}{\rule{0.900pt}{0.108pt}}
\multiput(779.00,402.17)(4.132,3.000){2}{\rule{0.450pt}{0.400pt}}
\multiput(785.00,406.61)(1.132,0.447){3}{\rule{0.900pt}{0.108pt}}
\multiput(785.00,405.17)(4.132,3.000){2}{\rule{0.450pt}{0.400pt}}
\multiput(791.00,409.61)(1.132,0.447){3}{\rule{0.900pt}{0.108pt}}
\multiput(791.00,408.17)(4.132,3.000){2}{\rule{0.450pt}{0.400pt}}
\multiput(797.00,412.61)(1.355,0.447){3}{\rule{1.033pt}{0.108pt}}
\multiput(797.00,411.17)(4.855,3.000){2}{\rule{0.517pt}{0.400pt}}
\multiput(804.00,415.61)(1.132,0.447){3}{\rule{0.900pt}{0.108pt}}
\multiput(804.00,414.17)(4.132,3.000){2}{\rule{0.450pt}{0.400pt}}
\multiput(810.00,418.61)(1.132,0.447){3}{\rule{0.900pt}{0.108pt}}
\multiput(810.00,417.17)(4.132,3.000){2}{\rule{0.450pt}{0.400pt}}
\multiput(816.00,421.61)(1.355,0.447){3}{\rule{1.033pt}{0.108pt}}
\multiput(816.00,420.17)(4.855,3.000){2}{\rule{0.517pt}{0.400pt}}
\multiput(823.00,424.61)(1.132,0.447){3}{\rule{0.900pt}{0.108pt}}
\multiput(823.00,423.17)(4.132,3.000){2}{\rule{0.450pt}{0.400pt}}
\multiput(829.00,427.61)(1.132,0.447){3}{\rule{0.900pt}{0.108pt}}
\multiput(829.00,426.17)(4.132,3.000){2}{\rule{0.450pt}{0.400pt}}
\multiput(835.00,430.61)(1.132,0.447){3}{\rule{0.900pt}{0.108pt}}
\multiput(835.00,429.17)(4.132,3.000){2}{\rule{0.450pt}{0.400pt}}
\put(841,433.17){\rule{1.500pt}{0.400pt}}
\multiput(841.00,432.17)(3.887,2.000){2}{\rule{0.750pt}{0.400pt}}
\multiput(848.00,435.61)(1.132,0.447){3}{\rule{0.900pt}{0.108pt}}
\multiput(848.00,434.17)(4.132,3.000){2}{\rule{0.450pt}{0.400pt}}
\multiput(854.00,438.61)(1.132,0.447){3}{\rule{0.900pt}{0.108pt}}
\multiput(854.00,437.17)(4.132,3.000){2}{\rule{0.450pt}{0.400pt}}
\multiput(860.00,441.61)(1.355,0.447){3}{\rule{1.033pt}{0.108pt}}
\multiput(860.00,440.17)(4.855,3.000){2}{\rule{0.517pt}{0.400pt}}
\multiput(867.00,444.61)(1.132,0.447){3}{\rule{0.900pt}{0.108pt}}
\multiput(867.00,443.17)(4.132,3.000){2}{\rule{0.450pt}{0.400pt}}
\multiput(873.00,447.61)(1.132,0.447){3}{\rule{0.900pt}{0.108pt}}
\multiput(873.00,446.17)(4.132,3.000){2}{\rule{0.450pt}{0.400pt}}
\multiput(879.00,450.61)(1.132,0.447){3}{\rule{0.900pt}{0.108pt}}
\multiput(879.00,449.17)(4.132,3.000){2}{\rule{0.450pt}{0.400pt}}
\multiput(885.00,453.61)(1.355,0.447){3}{\rule{1.033pt}{0.108pt}}
\multiput(885.00,452.17)(4.855,3.000){2}{\rule{0.517pt}{0.400pt}}
\multiput(892.00,456.61)(1.132,0.447){3}{\rule{0.900pt}{0.108pt}}
\multiput(892.00,455.17)(4.132,3.000){2}{\rule{0.450pt}{0.400pt}}
\multiput(898.00,459.61)(1.132,0.447){3}{\rule{0.900pt}{0.108pt}}
\multiput(898.00,458.17)(4.132,3.000){2}{\rule{0.450pt}{0.400pt}}
\multiput(904.00,462.61)(1.355,0.447){3}{\rule{1.033pt}{0.108pt}}
\multiput(904.00,461.17)(4.855,3.000){2}{\rule{0.517pt}{0.400pt}}
\multiput(911.00,465.61)(1.132,0.447){3}{\rule{0.900pt}{0.108pt}}
\multiput(911.00,464.17)(4.132,3.000){2}{\rule{0.450pt}{0.400pt}}
\multiput(917.00,468.61)(1.132,0.447){3}{\rule{0.900pt}{0.108pt}}
\multiput(917.00,467.17)(4.132,3.000){2}{\rule{0.450pt}{0.400pt}}
\multiput(923.00,471.61)(1.355,0.447){3}{\rule{1.033pt}{0.108pt}}
\multiput(923.00,470.17)(4.855,3.000){2}{\rule{0.517pt}{0.400pt}}
\multiput(930.00,474.61)(1.132,0.447){3}{\rule{0.900pt}{0.108pt}}
\multiput(930.00,473.17)(4.132,3.000){2}{\rule{0.450pt}{0.400pt}}
\multiput(936.00,477.61)(1.132,0.447){3}{\rule{0.900pt}{0.108pt}}
\multiput(936.00,476.17)(4.132,3.000){2}{\rule{0.450pt}{0.400pt}}
\multiput(942.00,480.61)(1.132,0.447){3}{\rule{0.900pt}{0.108pt}}
\multiput(942.00,479.17)(4.132,3.000){2}{\rule{0.450pt}{0.400pt}}
\multiput(948.00,483.61)(1.355,0.447){3}{\rule{1.033pt}{0.108pt}}
\multiput(948.00,482.17)(4.855,3.000){2}{\rule{0.517pt}{0.400pt}}
\multiput(955.00,486.61)(1.132,0.447){3}{\rule{0.900pt}{0.108pt}}
\multiput(955.00,485.17)(4.132,3.000){2}{\rule{0.450pt}{0.400pt}}
\put(961,489.17){\rule{1.300pt}{0.400pt}}
\multiput(961.00,488.17)(3.302,2.000){2}{\rule{0.650pt}{0.400pt}}
\multiput(967.00,491.61)(1.355,0.447){3}{\rule{1.033pt}{0.108pt}}
\multiput(967.00,490.17)(4.855,3.000){2}{\rule{0.517pt}{0.400pt}}
\multiput(974.00,494.61)(1.132,0.447){3}{\rule{0.900pt}{0.108pt}}
\multiput(974.00,493.17)(4.132,3.000){2}{\rule{0.450pt}{0.400pt}}
\multiput(980.00,497.61)(1.132,0.447){3}{\rule{0.900pt}{0.108pt}}
\multiput(980.00,496.17)(4.132,3.000){2}{\rule{0.450pt}{0.400pt}}
\multiput(986.00,500.61)(1.132,0.447){3}{\rule{0.900pt}{0.108pt}}
\multiput(986.00,499.17)(4.132,3.000){2}{\rule{0.450pt}{0.400pt}}
\multiput(992.00,503.61)(1.355,0.447){3}{\rule{1.033pt}{0.108pt}}
\multiput(992.00,502.17)(4.855,3.000){2}{\rule{0.517pt}{0.400pt}}
\multiput(999.00,506.61)(1.132,0.447){3}{\rule{0.900pt}{0.108pt}}
\multiput(999.00,505.17)(4.132,3.000){2}{\rule{0.450pt}{0.400pt}}
\multiput(1005.00,509.61)(1.132,0.447){3}{\rule{0.900pt}{0.108pt}}
\multiput(1005.00,508.17)(4.132,3.000){2}{\rule{0.450pt}{0.400pt}}
\multiput(1011.00,512.61)(1.355,0.447){3}{\rule{1.033pt}{0.108pt}}
\multiput(1011.00,511.17)(4.855,3.000){2}{\rule{0.517pt}{0.400pt}}
\multiput(1018.00,515.61)(1.132,0.447){3}{\rule{0.900pt}{0.108pt}}
\multiput(1018.00,514.17)(4.132,3.000){2}{\rule{0.450pt}{0.400pt}}
\multiput(1024.00,518.61)(1.132,0.447){3}{\rule{0.900pt}{0.108pt}}
\multiput(1024.00,517.17)(4.132,3.000){2}{\rule{0.450pt}{0.400pt}}
\multiput(1030.00,521.61)(1.132,0.447){3}{\rule{0.900pt}{0.108pt}}
\multiput(1030.00,520.17)(4.132,3.000){2}{\rule{0.450pt}{0.400pt}}
\multiput(1036.00,524.61)(1.355,0.447){3}{\rule{1.033pt}{0.108pt}}
\multiput(1036.00,523.17)(4.855,3.000){2}{\rule{0.517pt}{0.400pt}}
\multiput(1043.00,527.61)(1.132,0.447){3}{\rule{0.900pt}{0.108pt}}
\multiput(1043.00,526.17)(4.132,3.000){2}{\rule{0.450pt}{0.400pt}}
\multiput(1049.00,530.61)(1.132,0.447){3}{\rule{0.900pt}{0.108pt}}
\multiput(1049.00,529.17)(4.132,3.000){2}{\rule{0.450pt}{0.400pt}}
\multiput(1055.00,533.61)(1.355,0.447){3}{\rule{1.033pt}{0.108pt}}
\multiput(1055.00,532.17)(4.855,3.000){2}{\rule{0.517pt}{0.400pt}}
\multiput(1062.00,536.61)(1.132,0.447){3}{\rule{0.900pt}{0.108pt}}
\multiput(1062.00,535.17)(4.132,3.000){2}{\rule{0.450pt}{0.400pt}}
\multiput(1068.00,539.61)(1.132,0.447){3}{\rule{0.900pt}{0.108pt}}
\multiput(1068.00,538.17)(4.132,3.000){2}{\rule{0.450pt}{0.400pt}}
\multiput(1074.00,542.61)(1.132,0.447){3}{\rule{0.900pt}{0.108pt}}
\multiput(1074.00,541.17)(4.132,3.000){2}{\rule{0.450pt}{0.400pt}}
\multiput(1080.00,545.61)(1.355,0.447){3}{\rule{1.033pt}{0.108pt}}
\multiput(1080.00,544.17)(4.855,3.000){2}{\rule{0.517pt}{0.400pt}}
\put(1087,548.17){\rule{1.300pt}{0.400pt}}
\multiput(1087.00,547.17)(3.302,2.000){2}{\rule{0.650pt}{0.400pt}}
\multiput(1093.00,550.61)(1.132,0.447){3}{\rule{0.900pt}{0.108pt}}
\multiput(1093.00,549.17)(4.132,3.000){2}{\rule{0.450pt}{0.400pt}}
\multiput(1099.00,553.61)(1.355,0.447){3}{\rule{1.033pt}{0.108pt}}
\multiput(1099.00,552.17)(4.855,3.000){2}{\rule{0.517pt}{0.400pt}}
\multiput(1106.00,556.61)(1.132,0.447){3}{\rule{0.900pt}{0.108pt}}
\multiput(1106.00,555.17)(4.132,3.000){2}{\rule{0.450pt}{0.400pt}}
\multiput(1112.00,559.61)(1.132,0.447){3}{\rule{0.900pt}{0.108pt}}
\multiput(1112.00,558.17)(4.132,3.000){2}{\rule{0.450pt}{0.400pt}}
\multiput(1118.00,562.61)(1.132,0.447){3}{\rule{0.900pt}{0.108pt}}
\multiput(1118.00,561.17)(4.132,3.000){2}{\rule{0.450pt}{0.400pt}}
\multiput(1124.00,565.61)(1.355,0.447){3}{\rule{1.033pt}{0.108pt}}
\multiput(1124.00,564.17)(4.855,3.000){2}{\rule{0.517pt}{0.400pt}}
\multiput(1131.00,568.61)(1.132,0.447){3}{\rule{0.900pt}{0.108pt}}
\multiput(1131.00,567.17)(4.132,3.000){2}{\rule{0.450pt}{0.400pt}}
\multiput(1137.00,571.61)(1.132,0.447){3}{\rule{0.900pt}{0.108pt}}
\multiput(1137.00,570.17)(4.132,3.000){2}{\rule{0.450pt}{0.400pt}}
\multiput(1143.00,574.61)(1.355,0.447){3}{\rule{1.033pt}{0.108pt}}
\multiput(1143.00,573.17)(4.855,3.000){2}{\rule{0.517pt}{0.400pt}}
\multiput(1150.00,577.61)(1.132,0.447){3}{\rule{0.900pt}{0.108pt}}
\multiput(1150.00,576.17)(4.132,3.000){2}{\rule{0.450pt}{0.400pt}}
\multiput(1156.00,580.61)(1.132,0.447){3}{\rule{0.900pt}{0.108pt}}
\multiput(1156.00,579.17)(4.132,3.000){2}{\rule{0.450pt}{0.400pt}}
\multiput(1162.00,583.61)(1.355,0.447){3}{\rule{1.033pt}{0.108pt}}
\multiput(1162.00,582.17)(4.855,3.000){2}{\rule{0.517pt}{0.400pt}}
\multiput(1169.00,586.61)(1.132,0.447){3}{\rule{0.900pt}{0.108pt}}
\multiput(1169.00,585.17)(4.132,3.000){2}{\rule{0.450pt}{0.400pt}}
\multiput(1175.00,589.61)(1.132,0.447){3}{\rule{0.900pt}{0.108pt}}
\multiput(1175.00,588.17)(4.132,3.000){2}{\rule{0.450pt}{0.400pt}}
\multiput(1181.00,592.61)(1.132,0.447){3}{\rule{0.900pt}{0.108pt}}
\multiput(1181.00,591.17)(4.132,3.000){2}{\rule{0.450pt}{0.400pt}}
\multiput(1187.00,595.61)(1.355,0.447){3}{\rule{1.033pt}{0.108pt}}
\multiput(1187.00,594.17)(4.855,3.000){2}{\rule{0.517pt}{0.400pt}}
\multiput(1194.00,598.61)(1.132,0.447){3}{\rule{0.900pt}{0.108pt}}
\multiput(1194.00,597.17)(4.132,3.000){2}{\rule{0.450pt}{0.400pt}}
\multiput(1200.00,601.61)(1.132,0.447){3}{\rule{0.900pt}{0.108pt}}
\multiput(1200.00,600.17)(4.132,3.000){2}{\rule{0.450pt}{0.400pt}}
\put(1206,604.17){\rule{1.500pt}{0.400pt}}
\multiput(1206.00,603.17)(3.887,2.000){2}{\rule{0.750pt}{0.400pt}}
\multiput(1213.00,606.61)(1.132,0.447){3}{\rule{0.900pt}{0.108pt}}
\multiput(1213.00,605.17)(4.132,3.000){2}{\rule{0.450pt}{0.400pt}}
\multiput(1219.00,609.61)(1.132,0.447){3}{\rule{0.900pt}{0.108pt}}
\multiput(1219.00,608.17)(4.132,3.000){2}{\rule{0.450pt}{0.400pt}}
\multiput(1225.00,612.61)(1.132,0.447){3}{\rule{0.900pt}{0.108pt}}
\multiput(1225.00,611.17)(4.132,3.000){2}{\rule{0.450pt}{0.400pt}}
\multiput(1231.00,615.61)(1.355,0.447){3}{\rule{1.033pt}{0.108pt}}
\multiput(1231.00,614.17)(4.855,3.000){2}{\rule{0.517pt}{0.400pt}}
\multiput(1238.00,618.61)(1.132,0.447){3}{\rule{0.900pt}{0.108pt}}
\multiput(1238.00,617.17)(4.132,3.000){2}{\rule{0.450pt}{0.400pt}}
\multiput(1244.00,621.61)(1.132,0.447){3}{\rule{0.900pt}{0.108pt}}
\multiput(1244.00,620.17)(4.132,3.000){2}{\rule{0.450pt}{0.400pt}}
\multiput(1250.00,624.61)(1.355,0.447){3}{\rule{1.033pt}{0.108pt}}
\multiput(1250.00,623.17)(4.855,3.000){2}{\rule{0.517pt}{0.400pt}}
\multiput(1257.00,627.61)(1.132,0.447){3}{\rule{0.900pt}{0.108pt}}
\multiput(1257.00,626.17)(4.132,3.000){2}{\rule{0.450pt}{0.400pt}}
\multiput(1263.00,630.61)(1.132,0.447){3}{\rule{0.900pt}{0.108pt}}
\multiput(1263.00,629.17)(4.132,3.000){2}{\rule{0.450pt}{0.400pt}}
\multiput(1269.00,633.61)(1.132,0.447){3}{\rule{0.900pt}{0.108pt}}
\multiput(1269.00,632.17)(4.132,3.000){2}{\rule{0.450pt}{0.400pt}}
\multiput(1275.00,636.61)(1.355,0.447){3}{\rule{1.033pt}{0.108pt}}
\multiput(1275.00,635.17)(4.855,3.000){2}{\rule{0.517pt}{0.400pt}}
\multiput(1282.00,639.61)(1.132,0.447){3}{\rule{0.900pt}{0.108pt}}
\multiput(1282.00,638.17)(4.132,3.000){2}{\rule{0.450pt}{0.400pt}}
\multiput(1288.00,642.61)(1.132,0.447){3}{\rule{0.900pt}{0.108pt}}
\multiput(1288.00,641.17)(4.132,3.000){2}{\rule{0.450pt}{0.400pt}}
\multiput(1294.00,645.61)(1.355,0.447){3}{\rule{1.033pt}{0.108pt}}
\multiput(1294.00,644.17)(4.855,3.000){2}{\rule{0.517pt}{0.400pt}}
\multiput(1301.00,648.61)(1.132,0.447){3}{\rule{0.900pt}{0.108pt}}
\multiput(1301.00,647.17)(4.132,3.000){2}{\rule{0.450pt}{0.400pt}}
\multiput(1307.00,651.61)(1.132,0.447){3}{\rule{0.900pt}{0.108pt}}
\multiput(1307.00,650.17)(4.132,3.000){2}{\rule{0.450pt}{0.400pt}}
\multiput(1313.00,654.61)(1.132,0.447){3}{\rule{0.900pt}{0.108pt}}
\multiput(1313.00,653.17)(4.132,3.000){2}{\rule{0.450pt}{0.400pt}}
\multiput(1319.00,657.61)(1.355,0.447){3}{\rule{1.033pt}{0.108pt}}
\multiput(1319.00,656.17)(4.855,3.000){2}{\rule{0.517pt}{0.400pt}}
\put(1326,660.17){\rule{1.300pt}{0.400pt}}
\multiput(1326.00,659.17)(3.302,2.000){2}{\rule{0.650pt}{0.400pt}}
\multiput(1332.00,662.61)(1.132,0.447){3}{\rule{0.900pt}{0.108pt}}
\multiput(1332.00,661.17)(4.132,3.000){2}{\rule{0.450pt}{0.400pt}}
\multiput(1338.00,665.61)(1.355,0.447){3}{\rule{1.033pt}{0.108pt}}
\multiput(1338.00,664.17)(4.855,3.000){2}{\rule{0.517pt}{0.400pt}}
\multiput(1345.00,668.61)(1.132,0.447){3}{\rule{0.900pt}{0.108pt}}
\multiput(1345.00,667.17)(4.132,3.000){2}{\rule{0.450pt}{0.400pt}}
\multiput(1351.00,671.61)(1.132,0.447){3}{\rule{0.900pt}{0.108pt}}
\multiput(1351.00,670.17)(4.132,3.000){2}{\rule{0.450pt}{0.400pt}}
\multiput(1357.00,674.61)(1.355,0.447){3}{\rule{1.033pt}{0.108pt}}
\multiput(1357.00,673.17)(4.855,3.000){2}{\rule{0.517pt}{0.400pt}}
\multiput(1364.00,677.61)(1.132,0.447){3}{\rule{0.900pt}{0.108pt}}
\multiput(1364.00,676.17)(4.132,3.000){2}{\rule{0.450pt}{0.400pt}}
\multiput(1370.00,680.61)(1.132,0.447){3}{\rule{0.900pt}{0.108pt}}
\multiput(1370.00,679.17)(4.132,3.000){2}{\rule{0.450pt}{0.400pt}}
\multiput(1376.00,683.61)(1.132,0.447){3}{\rule{0.900pt}{0.108pt}}
\multiput(1376.00,682.17)(4.132,3.000){2}{\rule{0.450pt}{0.400pt}}
\multiput(1382.00,686.61)(1.355,0.447){3}{\rule{1.033pt}{0.108pt}}
\multiput(1382.00,685.17)(4.855,3.000){2}{\rule{0.517pt}{0.400pt}}
\multiput(1389.00,689.61)(1.132,0.447){3}{\rule{0.900pt}{0.108pt}}
\multiput(1389.00,688.17)(4.132,3.000){2}{\rule{0.450pt}{0.400pt}}
\multiput(1395.00,692.61)(1.132,0.447){3}{\rule{0.900pt}{0.108pt}}
\multiput(1395.00,691.17)(4.132,3.000){2}{\rule{0.450pt}{0.400pt}}
\multiput(1401.00,695.61)(1.355,0.447){3}{\rule{1.033pt}{0.108pt}}
\multiput(1401.00,694.17)(4.855,3.000){2}{\rule{0.517pt}{0.400pt}}
\multiput(1408.00,698.61)(1.132,0.447){3}{\rule{0.900pt}{0.108pt}}
\multiput(1408.00,697.17)(4.132,3.000){2}{\rule{0.450pt}{0.400pt}}
\multiput(1414.00,701.61)(1.132,0.447){3}{\rule{0.900pt}{0.108pt}}
\multiput(1414.00,700.17)(4.132,3.000){2}{\rule{0.450pt}{0.400pt}}
\multiput(1420.00,704.61)(1.132,0.447){3}{\rule{0.900pt}{0.108pt}}
\multiput(1420.00,703.17)(4.132,3.000){2}{\rule{0.450pt}{0.400pt}}
\multiput(1426.00,707.61)(1.355,0.447){3}{\rule{1.033pt}{0.108pt}}
\multiput(1426.00,706.17)(4.855,3.000){2}{\rule{0.517pt}{0.400pt}}
\multiput(1433.00,710.61)(1.132,0.447){3}{\rule{0.900pt}{0.108pt}}
\multiput(1433.00,709.17)(4.132,3.000){2}{\rule{0.450pt}{0.400pt}}
\end{picture}